\documentclass[aps,nofootinbib,superscriptaddress]{revtex4}%
\usepackage{hyperref}
\usepackage{amsmath}
\usepackage{amsfonts}
\usepackage{amssymb}
\usepackage{graphicx}
\usepackage{color}%
\providecommand{\U}[1]{\protect\rule{.1in}{.1in}}
\begin{document}
\title{Stringy scaling of $n$-point Regge string scattering amplitudes}
\author{Sheng-Hong Lai}
\email{xgcj944137@gmail.com}
\affiliation{Department of Electrophysics, National Yang Ming Chiao Tung University,
Hsinchu, Taiwan, R.O.C.}
\author{Jen-Chi Lee}
\email{jcclee@cc.nctu.edu.tw}
\affiliation{Department of Electrophysics, National Yang Ming Chiao Tung University,
Hsinchu, Taiwan, R.O.C.}
\affiliation{Institute of Physics, University of Tokyo, Komaba, Meguro-ku, Tokyo 153-8902, Japan}
\author{Yi Yang}
\email{yiyang@mail.nctu.edu.tw}
\affiliation{Department of Electrophysics, National Yang Ming Chiao Tung University,
Hsinchu, Taiwan, R.O.C.}
\date{\today}

\begin{abstract}
We discover a \textit{stringy scaling} behavior for a class of $n$-point Regge
string scattering amplitudes (RSSA). The number of independent kinematics
variables is found to be reduced by dim$\mathcal{M}$.

\end{abstract}
\maketitle

\bigskip%
%TCIMACRO{\TeXButton{equation number}{\setcounter{equation}{0}
%\renewcommand{\theequation}{\arabic{section}.\arabic{equation}}}}%
%BeginExpansion
\setcounter{equation}{0}
\renewcommand{\theequation}{\arabic{section}.\arabic{equation}}%
%EndExpansion

\section{Introduction}

Recent development of string scattering amplitudes (SSA) has shown that a
class of $4$-point SSA form representations of the $SL(K+3,C)$ group
\cite{Group,LSSA}. These are SSA with three tachyons and one arbitrary string
states%
\begin{equation}
\left\vert r_{n}^{T},r_{m}^{P},r_{l}^{L}\right\rangle =\prod_{n>0}\left(
\alpha_{-n}^{T}\right)  ^{r_{n}^{T}}\prod_{m>0}\left(  \alpha_{-m}^{P}\right)
^{r_{m}^{P}}\prod_{l>0}\left(  \alpha_{-l}^{L}\right)  ^{r_{l}^{L}}%
|0,k\rangle\label{state}%
\end{equation}
where $e^{P}=\frac{1}{M_{2}}(E_{2},\mathrm{k}_{2},0)=\frac{k_{2}}{M_{2}}$ is
the momentum polarization, $e^{L}=\frac{1}{M_{2}}(\mathrm{k}_{2},E_{2},0)$ is
the longitudinal polarization and $e^{T}=(0,0,1)$ is the transverse
polarization on the $\left(  2+1\right)  $-dimensional scattering plane. Note
that SSA of three tachyons and one arbitrary string states with polarizations
orthogonal to the scattering plane vanish. In addition to the mass level
$M_{2}^{2}=2(N-1)$ with%
\begin{equation}
N=\sum_{\substack{n,m,l>0\\\{\text{ }r_{j}^{X}\neq0\}}}\left(  nr_{n}%
^{T}+mr_{m}^{P}+lr_{l}^{L}\right)  , \label{NN}%
\end{equation}
another important index $K$ was identified for the state in Eq.(\ref{state})
\cite{LLY2}%
\begin{equation}
K=\sum_{\substack{n,m,l>0\\\{\text{ }r_{j}^{X}\neq0\}}}\left(  n+m+l\right)
\label{KKK}%
\end{equation}
where $X=\left(  T,P,L\right)  $ and one has put $r_{n}^{T}=r_{m}^{P}%
=r_{l}^{L}=1$ in Eq.(\ref{NN}) in the definition of $K$ in Eq.(\ref{KKK}).
Intuitively, $K$ counts the number of variety of the $\alpha_{-j}^{X}$
oscillators in Eq.(\ref{state}).

The representation bases of the above subclass of $4$-point SSA was soon
extended to all $4$-point SSA with arbitrary four string states, and
eventually to all $n$-point SSA with arbitrary $n$ string states
\cite{LLY3,exact}. It is thus important to study whether other known
interesting characteristics of the $4$-point SSA can be similarly extended to
the $n$-point SSA. One such characteristics of the $4$-point SSA is the
existence of infinite linear relations and their associated \textit{constant
ratios}, independent of the scattering angle $\phi$, among hard SSA (HSSA) at
each fixed mass level of the open bosonic string spectrum. These infinite
linear relations and their associated constant ratios were first conjectured
by Gross \cite{Gross,Gross1} and later explicitly calculated by the method of
decoupling of zero-norm states in \cite{ChanLee,ChanLee2,CHL,CHLTY2,CHLTY1}.

Indeed, in one of the authors' previous publications \cite{NHard}, we
discovered a general \textit{stringy scaling} behavior for all $n$-point HSSA
to all string loop orders. For the simplest case of $n=4$, the stringy scaling
behavior reduces to the infinite linear relations and the constant ratios of
HSSA at each mass level mentioned above. For this case, the ratios are
independent of $1$ scattering angle $\phi$ and thus the number of independent
kinematics variable reduced from $1$ to $0$ with dim$\mathcal{M=}$ $1-0=1$. In
general higher $n$-point HSSA, the stringy scaling behavior implies that the
number of independent kinematics variables of the ratios reduced by
dim$\mathcal{M}$ \cite{NHard}. See the definition of dim$\mathcal{M}$ in
Eq.(\ref{dim}) and Eq.(\ref{dimm}). As a result, the linear relations and
their associated constant ratios of $4$-point HSSA persist only in the
parameter spaces $\mathcal{M}$ for the cases of higher $n$-point HSSA
\cite{NHard}. See the example of constant ratios calculated among\textbf{ }%
$6$-point HSSA in Eq.(\ref{6ratio}).

In this paper, we will extend our calculation of stringy scaling behavior of
HSSA to the case of Regge SSA (RSSA). We will demonstrate a stringy scaling
behavior for a class of $n$-point RSSA, and the number of independent
kinematics variables is again found to be reduced by dim$\mathcal{M}$.

This paper is organized as following. In section II, we review and give a
detailed calculation of the stringy scaling behavior of HSSA \cite{NHard}. In
section III, we give a saddle point calculation of the hard stringy scaling to
justify the zero norm state (ZNS) calculation in section II. Section IV and V
are the main parts of this paper and we extend the calculation of HSSA to the
stringy scaling of RSSA. We will derive a stringy scaling behavior for a class
of $n$-point RSSA with arbitrary $n$ in section V. A brief conclusion was
given in section VI.%

%TCIMACRO{\TeXButton{equation number}{\setcounter{equation}{0}
%\renewcommand{\theequation}{\arabic{section}.\arabic{equation}}}}%
%BeginExpansion
\setcounter{equation}{0}
\renewcommand{\theequation}{\arabic{section}.\arabic{equation}}%
%EndExpansion

\section{The hard stringy scaling}

A brief report on stringy scaling of $n$-point hard string scattering
amplitudes (HSSA) was recently given in \cite{NHard}. In this section, we will
first give a detailed calculation of hard string scaling behavior. This can
also be served as a preparartion for the calculation of stringy scaling of
Regge string scattering amplitudes (RSSA) to be discussed in section IV and V.

\subsection{Stringy scaling of $4$-point HSSA}

The first stringy scaling was conjectured by Gross in 1988 \cite{Gross} which
claimed that all $4$-point HSSA ($E\rightarrow\infty$, fixed $\phi$) at each
fixed mass level share the same functional form. That is, all HSSA at each
fixed mass level are proportional to each other with \textit{constant} ratios
independent of the scattering angle $\phi$.

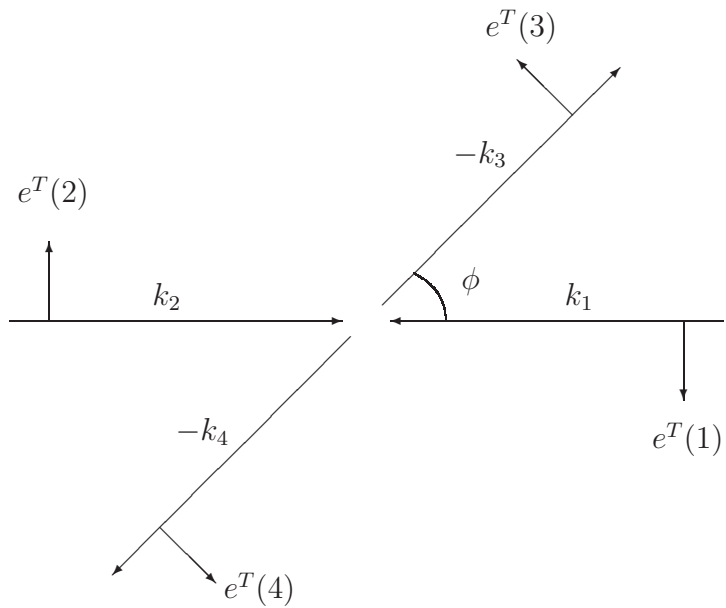
\begin{figure}[ptb]
\label{scattering} \setlength{\unitlength}{3pt}
\par
\begin{center}
\begin{picture}(100,70)(-50,-30)
{\large
% draw k_i
\put(45,0){\vector(-1,0){42}} \put(-45,0){\vector(1,0){42}}
\put(2,2){\vector(1,1){30}} \put(-2,-2){\vector(-1,-1){30}}
\put(25,2){$k_1$} \put(-27,2){$k_2$} \put(11,20){$-k_3$}
\put(-24,-15){$-k_4$}
% draw e^T^i
\put(40,0){\vector(0,-1){10}} \put(-40,0){\vector(0,1){10}}
\put(26,26){\vector(-1,1){7}} \put(-26,-26){\vector(1,-1){7}}
\put(36,-16){$e^{T}(1)$} \put(-44,15){$e^{T}(2)$}
\put(15,36){$e^{T}(3)$} \put(-18,-35){$e^{T}(4)$}
% draw \phi_{CM}
\qbezier(10,0)(10,4)(6,6) \put(12,4){$\phi$}
}
\end{picture}
\end{center}
\caption{Kinematic variables in the center of mass frame}%
\end{figure}

\bigskip To show this remarkable behavior, the starting point is to apply the
$4$-point $l$-loop stringy on-shell Ward identities \cite{ChanLee,ChanLee2}
\begin{equation}
\left\langle V_{1}\chi V_{3}V_{4}\right\rangle _{l-loop}=0 \label{vv}%
\end{equation}
in the hard scattering limit. In Eq.(\ref{vv}) $V_{j}$ above can be any string
vertex and the second vertex $\chi$ is the vertex of a zero-norm state (ZNS).
In the hard scattering limit, components of polarization orthogonal to the
scattering plane are subleading order in energy. On the other hand, it can be
shown that at each fixed mass level $M^{2}=2(N-1)$ only states of the
following form \cite{CHLTY2,CHLTY1} (in the hard scattering limit $e^{P}%
\simeq$ $e^{L}$)
\begin{equation}
\left\vert N,2m,q\right\rangle =\left(  \alpha_{-1}^{T}\right)  ^{N-2m-2q}%
\left(  \alpha_{-1}^{L}\right)  ^{2m}\left(  \alpha_{-2}^{L}\right)
^{q}\left\vert 0;k\right\rangle \label{11}%
\end{equation}
are leading order in energy.

There are two types of physical ZNS in the old covariant first quantized open
bosonic string spectrum: \cite{GSW}
\begin{equation}
\text{Type I}:L_{-1}\left\vert y\right\rangle ,\text{ where }L_{1}\left\vert
y\right\rangle =L_{2}\left\vert y\right\rangle =0,\text{ }L_{0}\left\vert
y\right\rangle =0;
\end{equation}%
\begin{equation}
\text{Type II}:(L_{-2}+\frac{3}{2}L_{-1}^{2})\left\vert \widetilde{y}%
\right\rangle ,\text{ where }L_{1}\left\vert \widetilde{y}\right\rangle
=L_{2}\left\vert \widetilde{y}\right\rangle =0,\text{ }(L_{0}+1)\left\vert
\widetilde{y}\right\rangle =0.\text{(}D=26\text{ \textbf{only}).}%
\end{equation}
\qquad

\textbf{(1)} We first consider $\chi$ to be the type I hard ZNS (HZNS)
calculated from Type I ZNS
\begin{align}
L_{-1}|N-1,2m-1,q\rangle &  =(M\alpha_{-1}^{L}+\alpha_{-2}^{L}\alpha_{1}%
^{L}+\underset{irrelevant}{\underbrace{\alpha_{-2}^{T}\alpha_{1}^{T}%
+\alpha_{-3}\cdot\alpha_{2}+\cdots}})|N-1,2m-1,q\rangle\nonumber\\
&  \simeq M|N,2m,q\rangle+(2m-1)|N,2m-2,q+1\rangle
\end{align}
where many terms are omitted because they are not of the form of
Eq.(\ref{11}). This implies the following relation among $4$-point amplitudes
\begin{equation}
\mathcal{T}^{(N,2m,q)}=-\frac{2m-1}{M}\mathcal{T}^{(N,2m-2,q+1)}.
\end{equation}
Using this relation repeatedly, we get
\begin{equation}
\mathcal{T}^{(N,2m,q)}=\frac{(2m-1)!!}{(-M)^{m}}\mathcal{T}^{(N,0,m+q)}.
\label{L1}%
\end{equation}

\textbf{(2)} Next, we consider another class of HZNS calculated from type II
ZNS
\begin{align}
L_{-2}|N-2,0,q\rangle &  =(\frac{1}{2}\alpha_{-1}^{T}\alpha_{-1}^{T}%
+M\alpha_{-2}^{L}+\underset{irrelevant}{\underbrace{\alpha_{-3}\cdot\alpha
_{1}+\cdots}})|N-2,0,q\rangle\nonumber\\
&  \simeq\frac{1}{2}|N,0,q\rangle+M|N,0,q+1\rangle.
\end{align}
Again, irrelevant terms are omitted here. From this we deduce that
\begin{equation}
\mathcal{T}^{(N,0,q+1)}=-\frac{1}{2M}\mathcal{T}^{(N,0,q)},
\end{equation}
which leads to
\begin{equation}
\mathcal{T}^{(N,0,q)}=\frac{1}{(-2M)^{q}}\mathcal{T}^{(N,0,0)}. \label{L2}%
\end{equation}

In conclusion, the decoupling of ZNS in Eq.(\ref{L1}) and Eq.(\ref{L2}) leads
to\textbf{ }constant ratios among $4$-point HSSA
\cite{ChanLee,ChanLee2,CHLTY2,CHLTY1}%

\begin{equation}
\frac{\mathcal{T}^{\left(  N,2m,q\right)  }}{\mathcal{T}^{\left(
N,0,0\right)  }}=\frac{\left(  2m\right)  !}{m!}\left(  \frac{-1}{2M}\right)
^{2m+q}.\text{(\textbf{independent of }}\phi\text{ !)} \label{22}%
\end{equation}
In Eq.(\ref{22}) $\mathcal{T}^{\left(  N,2m,q\right)  }$ is the $4$-point HSSA
of any string vertex $V_{j}$ with $j=1,3,4$ and $V_{2}$ is the high energy
state in Eq.(\ref{11}); while $\mathcal{T}^{\left(  N,0,0\right)  }$ is the
$4$-point HSSA of any string vertex $V_{j}$ with $j=1,3,4$, and $V_{2}$ is the
leading Regge trajectory string state at mass level $N$. Note that we have
omitted the tensor indice of $V_{j}$ with $j=1,3,4$ and keep only those of
$V_{2}$ in $\mathcal{T}^{\left(  N,2m,q\right)  }$.

\subsection{Examples of $4$-point stringy scaling}

\subsubsection{Bosonic open string}

Since the ratios of the amplitudes in Eq.(\ref{22}) are independent of the
choices of $V_{1}$, $V_{3}$ and $V_{4}$, we choose them to be tachyons and
$V_{2}$ to be Eq.(\ref{11}). On the other hand, since the ratios are
independent of the loop order, we choose to calculate HSSA of $l=0$ loop. An
explicit amplitude calculation for $M^{2}=4$, $6$ and $8$ gives
\cite{ChanLee,ChanLee2,CHLTY2,CHLTY1}%
\begin{equation}
\mathcal{T}_{TTT}:\mathcal{T}_{(LLT)}:\mathcal{T}_{(LT)}:\mathcal{T}%
_{[LT]}=8:1:-1:-1,
\end{equation}

\begin{align}
&  \mathcal{T}_{(TTTT)}:\mathcal{T}_{(TTLL)}:\mathcal{T}_{(LLLL)}%
:\mathcal{T}_{TT,L}:\mathcal{T}_{(TTL)}:\mathcal{T}_{(LLL)}:\mathcal{T}%
_{(LL)}\nonumber\\
&  =16:\frac{4}{3}:\frac{1}{3}:-\frac{2\sqrt{6}}{3}:-\frac{4\sqrt{6}}%
{9}:-\frac{\sqrt{6}}{9}:\frac{2}{3} \label{61}%
\end{align}
and%
\begin{align}
&  \mathcal{T}_{(TTTTT)}:\mathcal{T}_{(TTTL)}:\mathcal{T}_{(TTTLL)}%
:\mathcal{T}_{(TLLL)}:\mathcal{T}_{(TLLLL)}:\mathcal{T}_{(TLL)}:\mathcal{T}%
_{T,LL}:\mathcal{T}_{TLL,L}:\mathcal{T}_{TTT,L}\nonumber\\
&  =32:\sqrt{2}:2:\frac{3\sqrt{2}}{16}:\frac{3}{8}:\frac{1}{3}:\frac{2}%
{3}:\frac{\sqrt{2}}{16}:3\sqrt{2}, \label{mass8}%
\end{align}
respectively. These are all remarkably consistent with Eq.(\ref{22}) of ZNS
calculation \cite{review,over}.

It is important to note that for subleading order amplitudes, they are in
general \textit{not} proportional to each other. For $M^{2}=4$, for example,
one gets $6$ subleading order amplitudes and $4$ linear relations (on-shell
Ward identities) in the ZNS calculation. An explicit subleading order
amplitude calculation gives \cite{ChanLee,ChanLee2}%

\begin{align}
\mathcal{T}_{LLL}^{2}  &  \sim-4E^{8}\sin\phi\cos\phi,\nonumber\\
\mathcal{T}_{LTT}^{2}  &  \sim-8E^{8}\sin^{2}\phi\cos\phi,
\end{align}
which show that the proportional coefficients do depend on the scattering
angle $\phi$.

\subsubsection{Bosonic closed string and D-particle}

For closed string scatterings \cite{Closed,bosonic2}, one can use the KLT
formula \cite{KLT}, which expresses the relation between tree amplitudes of
closed and two channels of open string $(\alpha_{\text{closed}}^{\prime
}=4\alpha_{\text{open}}^{\prime}=2),$ to obtain the closed string ratios which
are the tensor product of two open string ratios in Eq.(\ref{22}). On the
other hand, it is interesting to find that the ratios of hard closed string
D-particle scatterings are again given by the tensor product of two open
string ratios \cite{LMY}%
\begin{equation}
\frac{T_{SD}^{\left(  N;2m,2m^{^{\prime}};q,q^{^{\prime}}\right)  }}%
{T_{SD}^{\left(  N;0,0;0,0\right)  }}=\left(  -\frac{1}{M_{2}}\right)
^{2(m+m^{^{\prime}})+q+q^{^{\prime}}}\left(  \frac{1}{2}\right)
^{m+m^{^{\prime}}+q+q^{^{\prime}}}(2m-1)!!(2m^{\prime}-1)!!,
\end{equation}
which came as a surprise since there is no physical picture for open string
D-particle tree scattering amplitudes and thus no factorization for closed
string D-particle scatterings into two channels of open string D-particle
scatterings, and hence no KLT-like formula there. However, these ratios are
consistent with the decoupling of high energy ZNS calculation.

\subsubsection{Stringy scaling of Superstring}

It turned out to be nontrivial to extend the linear relations and their
associated constant ratios of the HSSA of bosonic string to the case of $10D$
open superstring. First of all, in addition to the NS-sector, there are
massive fermionic states in the R-sector whose vertex operators are still
unknown except the leading Regge trajectory states in the spectrum
\cite{Osch}. So the only known complete vertex operators so far are those for
the mass level $M^{2}=2$ \cite{RRR} which contains no off leading massive
Regge trajectory fermionic string states.

Secondly, in the NS-sector of $M^{2}=2$ it was surprised to note that
\cite{New} there exists no "inter-particle gauge transformation" induced by
bosonic ZNS for the two positive-norm physical propagating states, the
symmetric spin three and the anti-symmetric spin two states. However, the
$4$-point HSSA among these two positive-norm states are still related and are
indeed again proportional to each others. Presumably, this is due to the
massive spacetime SUSY and the existence of spacetime massive fermion string
scattering amplitudes of the R-sector of the theory \cite{spin}.

Thirdly, it was noted that for the HSSA of the NS sector of superstring, there
existed leading order HSSA with polarizations orthogonal to the scattering
plane \cite{New}. This was due to the \textquotedblright worldsheet fermion
exchange\textquotedblright\ \cite{susy} in the correlation functions and was
argued to be related to the HSSA of massive spacetime fermion of R-sector of
the theory \cite{spin}.

The first calculation of the $4$-point superstringy scaling was performed for
the NS-sector of $10D$ open superstring theory. There are four classes of HSSA
of superstring which are all proportional to each other \cite{susy}
\begin{align}
\left\vert N,2m,q\right\rangle \otimes\left\vert b_{-\frac{3}{2}}%
^{P}\right\rangle  &  =\left(  -\frac{1}{2M_{2}}\right)  ^{q+m}\frac{\left(
2m-1\right)  !!}{\left(  -M_{2}\right)  ^{m}}\left\vert N,0,0\right\rangle
\otimes\left\vert b_{-\frac{3}{2}}^{P}\right\rangle ,\label{SS1}\\
\left\vert N+1,2m+1,q\right\rangle \otimes\left\vert b_{-\frac{1}{2}}%
^{P}\right\rangle  &  =\left(  -\frac{1}{2M_{2}}\right)  ^{q+m}\frac{\left(
2m+1\right)  !!}{\left(  -M_{2}\right)  ^{m+1}}\left\vert N,0,0\right\rangle
\otimes\left\vert b_{-\frac{3}{2}}^{P}\right\rangle ,\label{SS2}\\
\left\vert N+1,2m,q\right\rangle \otimes\left\vert b_{-\frac{1}{2}}%
^{T}\right\rangle  &  =\left(  -\frac{1}{2M_{2}}\right)  ^{q+m}\frac{\left(
2m-1\right)  !!}{\left(  -M_{2}\right)  ^{m-1}}\left\vert N,0,0\right\rangle
\otimes\left\vert b_{-\frac{3}{2}}^{P}\right\rangle ,\label{SS3}\\
\left\vert N-1,2m,q-1\right\rangle \otimes\left\vert b_{-\frac{1}{2}}%
^{T}b_{-\frac{1}{2}}^{P}b_{-\frac{3}{2}}^{P}\right\rangle  &  =\left(
-\frac{1}{2M_{2}}\right)  ^{q+m}\frac{\left(  2m-1\right)  !!}{\left(
-M_{2}\right)  ^{m}}\left\vert N,0,0\right\rangle \otimes\left\vert
b_{-\frac{3}{2}}^{P}\right\rangle . \label{SS4}%
\end{align}

Note that, in order to simplify the notation, we have only shown the second
state of the four point functions to represent the scattering amplitudes on
both sides of each equation above. Eqs.(\ref{SS1}) to (\ref{SS4}) are thus the
SUSY generalization of Eq.(\ref{22}) for the bosonic string.

Moreover, a recent calculation showed that \cite{spin} among $2^{4}\times
2^{4}=256$ $4$-point polarized fermion SSA (PFSSA) in the R-sector of
$M^{2}=2$ states, only $16$ of them are of leading order in energy and all of
them share the same functional form in the hard scattering limit. On the other
hand, the ratios of the \textit{complete} $4$-point HSSA in the NS sector of
mass level $M^{2}=2$ which include HSSA with polarizations orthogonal to the
sacttering plane are \cite{New}%
\begin{align}
\left\langle b_{\frac{-1}{2}}^{T},\alpha_{-1}^{T}b_{\frac{-1}{2}}%
^{T}\right\rangle  &  :\left\langle b_{\frac{-1}{2}}^{T},\left(  2b_{\frac
{-1}{2}}^{L}\alpha_{-1}^{L}-Mb_{\frac{-3}{2}}^{L}\right)  \right\rangle
:\left\langle b_{\frac{-1}{2}}^{T_{i}},\alpha_{-1}^{T}b_{\frac{-1}{2}}^{T_{j}%
}\right\rangle :\left\langle b_{\frac{-1}{2}}^{T_{k}},b_{\frac{-1}{2}}%
^{L}b_{\frac{-1}{2}}^{T}b_{\frac{-1}{2}}^{T_{l}}\right\rangle \label{susy1}\\
&  =-2k_{3}^{T}E^{2}:-2(\frac{2}{M^{2}}+1)k_{3}^{T}E^{2}:\delta_{ij}2k_{3}%
^{T}E^{2}:\delta_{lk}\frac{-2k_{3}^{T}E^{2}}{M}\nonumber\\
&  =1:2:-\delta_{ij}:\frac{\delta_{lk}}{\sqrt{2}}.\text{ \ }(\text{\ }%
i,j,k,l=3,4,5,...,9) \label{susy3}%
\end{align}
where we have, for simplicity, omitted the last two tachyon vertices in the
notation of each HSSA in Eq.(\ref{susy1}). In sum, in the NS sector one gets
$1+1+7+7=16$ HSSA in Eq.(\ref{susy1}). This result agrees with those of $16$
hard massive PFSSA in the R-sector calculated recently \cite{spin}.

\subsubsection{Field theory}

On the other hand, in field theory, as an example, the leading order process
of the elastic scattering of a spin-$\frac{1}{2}$ particle by a spin-$0$
particle such as $e^{-}\pi^{+}\longrightarrow e^{-}\pi^{+}$, the non-vanishing
amplitudes were shown to be \cite{JW}%
\begin{align}
\mathcal{T}\text{ }(e_{R}^{-}\pi^{+}  &  \longrightarrow e_{R}^{-}\pi
^{+})=\mathcal{T}\text{ }(e_{L}^{-}\pi^{+}\longrightarrow e_{L}^{-}\pi
^{+})\sim\text{ }\cos\frac{\phi}{2},\\
\mathcal{T}\text{ }(e_{R}^{-}\pi^{+}  &  \longrightarrow e_{L}^{-}\pi
^{+})=\mathcal{T}\text{ }(e_{L}^{-}\pi^{+}\longrightarrow e_{R}^{-}\pi
^{+})\sim\text{ }\sin\frac{\phi}{2},
\end{align}
which are \textit{not} proportional to each other. In QED, as another example,
for the leading order process of $e^{-}e^{+}\longrightarrow\mu^{-}\mu^{+}$,
there are $4$ non-vanishing among $16$ hard polarized amplitudes \cite{Peskin}%
\begin{align}
\mathcal{T}\text{ }(e_{R}^{-}e_{L}^{+}  &  \longrightarrow\mu_{R}^{-}\mu
_{L}^{+})=\mathcal{T}\text{ }(e_{L}^{-}e_{R}^{+}\longrightarrow\mu_{L}^{-}%
\mu_{R}^{+})\sim\text{ }(1+\cos\theta)=2\text{ }\cos^{2}\frac{\phi}{2},\\
\mathcal{T}\text{ }(e_{R}^{-}e_{L}^{+}  &  \longrightarrow\mu_{L}^{-}\mu
_{R}^{+})=\mathcal{T}\text{ }(e_{L}^{-}e_{R}^{+}\longrightarrow\mu_{R}^{-}%
\mu_{L}^{+})\sim\text{ }(1-\cos\theta)=2\text{ }\sin^{2}\frac{\phi}{2},
\end{align}
and they are \textit{not} all proportional to each other.

\subsection{Stringy scaling of higher point ($n\geq5$) HSSA}

It is tempted to extend the stringy scaling behavior of $4$-point SSA derived
in the previous subsection to the higher point SSA. The $n$-point stringy
on-shell Ward identities can be written as%
\begin{equation}
\left\langle V_{1}\chi V_{3}\cdots V_{n}\right\rangle _{l-loop}=0 \label{www}%
\end{equation}
where $\chi$ again is the vertex of a ZNS. We begin the discussion with a
simple kinematics regime on the scattering plane.

\subsubsection{On the scattering plane}

In the hard scattering limit on the scattering plane, the space part of
momenta $k_{j}$ ( $j=3,4,\cdots,n$) form a closed $1$-chain with $(n-2)$ sides
due to momentum conservation. It turned out that all the $4$-point calculation
in the previous subsection persist and one ends up with Eq.(\ref{22}) again
\cite{NHard}. However, while for $n=4$ the \textit{ratios} are independent of
$1$ scattering angle $\phi$, for $n=5$, the ratios are independent of $3$
kinematics variables ($2$ angles and $1$ fixed ratio of two infinite energies)
or, for simplicity, $3$ scattering "angles". For $n=6$, there are $5$
scattering "angles" etc..

\subsubsection{Out of the scattering plane}

The general high energy states at each fixed mass level $M^{2}=2(N-1)$ can be
written as \cite{NHard}%
\begin{equation}
\left\vert \left\{  p_{i}\right\}  ,2m,q\right\rangle =\left(  \alpha
_{-1}^{T_{1}}\right)  ^{N+p_{1}}\left(  \alpha_{-1}^{T_{2}}\right)  ^{p_{2}%
}\cdots\left(  \alpha_{-1}^{T_{r}}\right)  ^{p_{r}}\left(  \alpha_{-1}%
^{L}\right)  ^{2m}\left(  \alpha_{-2}^{L}\right)  ^{q}\left\vert
0;k\right\rangle \label{HES}%
\end{equation}
where $\sum_{i=1}^{r}p_{i}=-2(m+q)$ with\ $r\leq24$. In Eq.(\ref{HES}),
$T_{j}$ is the $j$th transverse direction orthogonal to $k_{2}$. For higher
dimensional scattering space, one generalizes the transverse polarization
$e^{T}=(0,0,1)$ to $e^{\hat{T}}=(0,0,\vec{\omega})$ where%
\begin{equation}
\omega_{i}=\cos\theta_{i}\prod\limits_{\sigma=1}^{i-1}\sin\theta_{\sigma
}\text{with }i=1,\cdots,r,\text{ }\theta_{r}=0\text{ }%
\end{equation}
are the solid angles in the transverse space spanned by $24$ transverse
directions $e^{T_{i}}$. Note that $\alpha_{-1}^{\hat{T}}=\alpha_{-1}\cdot
e^{\hat{T}}$ etc. With $\left(  \alpha_{-1}^{T_{i}}\right)  =\left(
\alpha_{-1}^{\hat{T}}\right)  \omega_{i}$, we easily obtain%
\begin{align}
&  \left(  \alpha_{-1}^{T_{1}}\right)  ^{N+p_{1}}\left(  \alpha_{-1}^{T_{2}%
}\right)  ^{p_{2}}\cdots\left(  \alpha_{-1}^{T_{r}}\right)  ^{p_{r}}\left(
\alpha_{-1}^{L}\right)  ^{2m}\left(  \alpha_{-2}^{L}\right)  ^{q}\left\vert
0;k\right\rangle \nonumber\\
&  =\left(  \omega_{1}^{N}\prod_{i=1}^{r}\omega_{i}^{p_{i}}\right)  \left(
\alpha_{-1}^{\hat{T}}\right)  ^{N-2m-2q}\left(  \alpha_{-1}^{L}\right)
^{2m}\left(  \alpha_{-2}^{L}\right)  ^{q}\left\vert 0;k\right\rangle ,
\end{align}
which leads to the ratios of $n$-point HSSA \cite{NHard}%
\begin{equation}
\frac{\mathcal{T}^{\left(  \left\{  p_{i}\right\}  ,2m,q\right)  }%
}{\mathcal{T}^{\left(  \left\{  0_{i}\right\}  ,0,0\right)  }}=\frac{\left(
2m\right)  !}{m!}\left(  \frac{-1}{2M}\right)  ^{2m+q}\prod_{i=1}^{r}%
\omega_{i}^{p_{i}} \label{100}%
\end{equation}
where $\mathcal{T}^{\left(  \left\{  0_{i}\right\}  ,0,0\right)  }$ is the
HSSA of leading Regge trajectory state at mass level $M^{2}=2(N-1)$. It is
important to note that the number of kinematics variables dependence in the
ratios of Eq.(\ref{100}) reduced. This stringy scaling behavior of $n$-point
($n\geq5$) HSSA is the generalization of that of $4$-point HSSA in
Eq.(\ref{22}). Since the result of ZNS calculation in Eq.(\ref{100}) is based
on the stringy Ward identity in Eq.(\ref{www}), The ratios calculated in
Eq.(\ref{100}) are valid to all string loop orders\textbf{.}

\subsection{Degree of Stringy Scaling}

We see in the previous section that for the simple case with $n=4$ and $r=1$,
one has two variables, $s$ and $t$ (or $E$, $\phi$). The ratios of all HSSA
are independent of the scattering angle $\phi$ and we will call the degree of
the scaling dim$\mathcal{M}=1$. The dependence of the number of kinematics
variable reduced from $1$ to $0$ and we have $1-0=$ dim$\mathcal{M}=1$. (see
the definition of $\mathcal{M}$ below)

For the general $n$-point HSSA with $r\leq24$, $d=r+2$, we have $k_{j}$ vector
with $j=1,\cdots,n$ and $k_{j}$ $\in R^{d-1,1}$. The number of kinematics
variables is $n\left(  d-1\right)  -\frac{d\left(  d+1\right)  }{2}$. Indeed,
as $p=E\rightarrow\infty$ , that implies $q_{j}\rightarrow\infty$ in the hard
limit, we define the $26$-dimensional momenta in the CM frame to be%
\begin{align}
k_{1}  &  =\left(  E,-E,0^{r}\right)  ,\nonumber\\
k_{2}  &  =\left(  E,+E,0^{r}\right)  ,\nonumber\\
&  \vdots\nonumber\\
k_{j}  &  =\left(  -q_{j},-q_{j}\Omega_{1}^{j},-q_{j}\Omega_{2}^{j}%
,\cdots,-q_{j}\Omega_{r}^{j},-q_{j}\Omega_{r+1}^{j}\right)  \label{k22}%
\end{align}
where $j=3,4,\cdots,n$, and%
\begin{equation}
\Omega_{i}^{j}=\cos\phi_{i}^{j}\prod\limits_{\sigma=1}^{i-1}\sin\phi_{\sigma
}^{j}\text{ with }\phi_{j-1}^{j}=0,\text{ }\phi_{i>r}^{j}=0\text{ and }%
r\leq\min\left\{  n-3,24\right\}  \label{k33}%
\end{equation}
are the solid angles in the $\left(  j-2\right)  $-dimensional spherical space
with $\sum_{i=1}^{j-2}\left(  \Omega_{i}^{j}\right)  ^{2}=1$. In
Eq.(\ref{k22}), $0^{r}$ denotes the $r$-dimensional null vector. The condition
$\phi_{j-1}^{j}=0$ in Eq.(\ref{k33}) was chosen to fix the frame by using the
rotational symmetry.

The independent kinematics variables can be chosen to be some $\varphi_{i}%
^{j}$ and some fixed ratios of infinite $q_{j}$. For the kinematics parameter
space $\mathcal{M}$ defined by \cite{NHard}
\begin{equation}
\omega_{j}\left(  \text{kinematics parameters with }E\rightarrow\infty\right)
=\text{fixed constant \ }(j=2,\cdots,r)\text{,}%
\end{equation}
we can count the dimension of $\mathcal{M}$ to be \cite{NHard}
\begin{equation}
\text{dim}\mathcal{M}\text{ }\mathcal{=}\text{ }n\left(  d-1\right)
-\frac{d\left(  d+1\right)  }{2}-1-\left(  r-1\right)  =\frac{\left(
r+1\right)  \left(  2n-r-6\right)  }{2} \label{dim}%
\end{equation}
where $r=d-2$ is the number of transverse directions $e^{T_{i}}$. In sum, the
ratios among $n$-point $HSSA$ with $r\leq24$ are constants and independent of
the scattering "angles" in the kinematic regime $\mathcal{M}$.

\subsubsection{Examples}

(1). For $n=5$ and $r=2$, $d=r+2=4$ and one has $n\left(  d-1\right)
-\frac{d\left(  d+1\right)  }{2}=5$ parameters ($r_{1}$ is the ratio of two
infinite energies)%
\begin{equation}
E,\phi_{2}^{3},\phi_{2}^{4},\phi_{3}^{4},r_{1}.
\end{equation}
In the hard scattering limit $E\rightarrow\infty$, for $\theta_{1}=fixed$ we
get dim$\mathcal{M=}$ $3$.

(2). For $n=6$ and $r=3$, the ratios of $6$-point HSSA depends only on $2$
variables $\theta_{1}$ and $\theta_{2}$ instead of $8$ "angles" and
dim$\mathcal{M=}$ $6$. For this case, $\mathcal{M}$ is defined by
\begin{equation}
\theta_{j}\left(  8\text{ kinematics parameters}\right)  =\text{fixed
constant, \ }j=1,2,
\end{equation}
and the ratios \cite{NHard}%
\begin{equation}
\frac{\mathcal{T}^{\left(  \left\{  p_{1},p_{2},p_{3}\right\}  ,2m,q\right)
}}{\mathcal{T}^{\left(  \left\{  0,0,0\right\}  ,0,0\right)  }}=\frac{\left(
2m\right)  !}{m!}\left(  \frac{-1}{2M}\right)  ^{2m+q}\left(  \cos\theta
_{1}\right)  ^{p_{1}}\left(  \sin\theta_{1}\cos\theta_{2}\right)  ^{p_{2}%
}\left(  \sin\theta_{1}\sin\theta_{2}\right)  ^{p_{3}}%
\end{equation}
are independent of kinematics parameters in the space $\mathcal{M}$. For
example, for say $\theta_{1}=\frac{\pi}{4}$ and $\theta_{2}=\frac{\pi}{6}$, we
get the ratios among\textbf{ }$6$-point HSSA%
\begin{equation}
\frac{\mathcal{T}^{\left(  \left\{  p_{1},p_{2},p_{3}\right\}  ,2m,q\right)
}}{\mathcal{T}^{\left(  \left\{  0,0,0\right\}  ,0,0\right)  }}=\left(
-\frac{1}{M}\right)  ^{2m+q}(2m-1)!!\left(  \frac{1}{2}\right)  ^{p_{2}+p_{3}%
}\left(  \sqrt{3}\right)  ^{p_{3}}\text{.} \label{6ratio}%
\end{equation}
These ratios for higher point HSSA are one example of generalization of
previous ratios calculated in Eq.(\ref{22}) for the case $4$-point HSSA.

\subsubsection{General cases}

In general, in the hard scattering limit, the number of scattering "angles"
dependence on ratios of $n$-point $HSSA$ with $r\leq24$ reduces by
dim$\mathcal{M}$. For a given ($n,r$), we can calculate some examples of
dim$\mathcal{M}$ \cite{NHard}
\begin{equation}%
\begin{tabular}
[c]{lllll}%
$\text{dim}\mathcal{M}$ & $r=1$ & $r=2$ & $r=3$ & $r=4$\\
$n=4$ & $1$ &  &  & \\
$n=5$ & $3$ & $3$ &  & \\
$n=6$ & $5$ & $6$ & $6$ & \\
$n=7$ & $7$ & $9$ & $10$ & $10$%
\end{tabular}
\ \ \ \ \ \ \ \ \ \ \ \ \ \ . \label{dimm}%
\end{equation}
Note that for the $n=4$ and $r=1$ case, one obtains the previous $4$-point
case in Eq.(\ref{22}).%

%TCIMACRO{\TeXButton{equation number}{\setcounter{equation}{0}
%\renewcommand{\theequation}{\arabic{section}.\arabic{equation}}}}%
%BeginExpansion
\setcounter{equation}{0}
\renewcommand{\theequation}{\arabic{section}.\arabic{equation}}%
%EndExpansion

\section{Saddle point calculation}

To justify the ZNS calculation in Eq.(\ref{22}) and Eq.(\ref{100}), we use the
saddle point calculation to explicitly calculate the HSSA. Since the ratios
are independent of the choices of $V_{j}$ ($J=1,3,4\cdots,n$), we choose them
to be tachyons and $V_{2}$ to be the high energy state in Eq.(\ref{11}). On
the other hand, since the ratios are independent of the loop order, we choose
to calculate $l=0$ loop. We begin with the $4$-point case \cite{CHLTY2,CHLTY1}.

\subsection{The four point calculation}

The $t-u$ channel contribution to the stringy amplitude at tree level is
(after $SL(2,R)$ fixing)%
\begin{align}
\mathcal{T}^{(N,2m,q)}  &  =\int_{1}^{\infty}dxx^{(1,2)}(1-x)^{(2,3)}\left[
\frac{e^{T}\cdot k_{1}}{x}-\frac{e^{T}\cdot k_{3}}{1-x}\right]  ^{N-2m-2q}%
\nonumber\\
&  \cdot\left[  \frac{e^{P}\cdot k_{1}}{x}-\frac{e^{P}\cdot k_{3}}%
{1-x}\right]  ^{2m}\left[  -\frac{e^{P}\cdot k_{1}}{x^{2}}-\frac{e^{P}\cdot
k_{3}}{(1-x)^{2}}\right]  ^{q}%
\end{align}
where $(1,2)=k_{1}\cdot k_{2}$ etc.

In order to apply the saddle-point method, we rewrite the amplitude above into
the following form%
\begin{equation}
\mathcal{T}^{(N,2m,q)}(K)=\int_{1}^{\infty}dx\mbox{ }u(x)e^{-Kf(x)},
\end{equation}
where
\begin{align}
K  &  \equiv-(1,2)\rightarrow\frac{s}{2}\rightarrow2E^{2},\\
\tau &  \equiv-\frac{(2,3)}{(1,2)}\rightarrow-\frac{t}{s}\rightarrow\sin
^{2}\frac{\phi}{2},\\
f(x)  &  \equiv\ln x-\tau\ln(1-x),\\
u(x)  &  \equiv\left[  \frac{(1,2)}{M}\right]  ^{2m+q}(1-x)^{-N+2m+2q}%
\underset{}{\underbrace{(f^{\prime})^{2m}}}(f^{\prime\prime})^{q}(-e^{T}\cdot
k_{3})^{N-2m-2q}. \label{factor}%
\end{align}
The saddle-point for the integration of moduli, $x=x_{0}$, is defined by
\begin{equation}
f^{\prime}(x_{0})=0,
\end{equation}
and we have%
\begin{equation}
x_{0}=\frac{1}{1-\tau}=\sec^{2}\frac{\phi}{2},\hspace{1cm}1-x_{0}=-\frac{\tau
}{1-\tau},\hspace{1cm}f^{\prime\prime}(x_{0})=(1-\tau)^{3}\tau^{-1}.
\label{saddle}%
\end{equation}
Due to the factor $(f^{\prime})^{2m}$ in Eq.(\ref{factor}), it is easy to see
that \cite{CHLTY2,CHLTY1}%
\begin{equation}
u(x_{0})=u^{\prime}(x_{0})=....=u^{(2m-1)}(x_{0})=0,
\end{equation}
and
\begin{equation}
u^{(2m)}(x_{0})=\left[  \frac{(1,2)}{M}\right]  ^{2m+q}(1-x_{0})^{-N+2m+2q}%
(2m)!(f_{0}^{\prime\prime})^{2m+q}(-e^{T}\cdot k_{3})^{N-2m-2q}.
\end{equation}

With these inputs, one can easily evaluate the Gaussian integral associated
with the four-point amplitudes \cite{CHLTY2,CHLTY1}
\begin{align}
&  \int_{1}^{\infty}dx\mbox{ }u(x)e^{-Kf(x)}\nonumber\\
&  =\sqrt{\frac{2\pi}{Kf_{0}^{\prime\prime}}}e^{-Kf_{0}}\left[  \frac
{u_{0}^{(2m)}}{2^{m}\ m!\ (f_{0}^{\prime\prime})^{m}\ K^{m}}+O(\frac
{1}{K^{m+1}})\right] \nonumber\\
&  =\sqrt{\frac{2\pi}{Kf_{0}^{\prime\prime}}}e^{-Kf_{0}}\left[  (-1)^{N-q}%
\frac{2^{N-2m-q}(2m)!}{m!\ {M}^{2m+q}}\ \tau^{-\frac{N}{2}}(1-\tau)^{\frac
{3N}{2}}E^{N}+O(E^{N-2})\right]  .
\end{align}
This result shows explicitly that with one tensor and three tachyons, the
energy and angle dependence for the four-point HSS amplitudes only depend on
the level $N$ \cite{CHLTY2,CHLTY1}
\begin{align}
\lim_{E\rightarrow\infty}\frac{\mathcal{T}^{(N,2m,q)}}{\mathcal{T}^{(N,0,0)}}
&  =\frac{(-1)^{q}(2m)!}{m!(2M)^{2m+q}}\nonumber\\
&  =(-\frac{2m-1}{M})....(-\frac{3}{M})(-\frac{1}{M})(-\frac{1}{2M})^{m+q},
\end{align}
which is\textbf{ }remarkably\textbf{ }consistent with calculation of
decoupling of high energy ZNS obtained in Eq.(\ref{22}).

\subsection{The $n$-point HSSA with $r=1$}

To illustrate the $n$-point HSSA calculation, we begin with $n$-point HSSA
with $r=1$. We want to calculate $n$-point HSSA with $(n-1)$ tachyons and $1$
high energy state in Eq.(\ref{11}). With the change of variables $z_{i}%
=\frac{x_{i}}{x_{i+1}}$ or $x_{i}=z_{i}\cdots z_{n-2}$, the HSSA can be
written as%
\begin{align}
\mathcal{T}^{\left(  \left\{  p_{i}\right\}  ,m,q\right)  }  &  =\int_{0}%
^{1}dx_{n-2}\cdots\text{ }\int_{0}^{x_{4}}dx_{3}\int_{0}^{x_{3}}dx_{2}%
ue^{-Kf}\nonumber\\
&  =\int_{0}^{1}dz_{n-2}\cdots\text{ }\int_{0}^{1}dz_{3}\int_{0}^{1}dz_{2}%
\begin{vmatrix}
z_{3}\cdots z_{n-2} & z_{2}z_{4}\cdots z_{n-2} & \cdots & z_{2}\cdots
z_{n-3}\\
0 & z_{4}\cdots z_{n-2} & \cdots & \\
&  & \ddots & \\
0 & 0 & \cdots & 1
\end{vmatrix}
ue^{-Kf}\nonumber\\
&  =\left(  \prod_{i=3}^{n-2}\int_{0}^{1}dz_{i}\text{ }z_{i}^{i-2-N}\right)
\int_{0}^{1}dz_{2}ue^{-Kf}%
\end{align}
where%
\begin{align}
f\left(  x_{i}\right)   &  =-\underset{i<j}{\sum}\frac{k_{i}\cdot k_{j}}{K}%
\ln\left(  x_{j}-x_{i}\right)  =-\underset{i<j}{\sum}\frac{k_{i}\cdot k_{j}%
}{K}\ln\left(  z_{j}\cdots z_{n-2}-z_{i}\cdots z_{n-2}\right) \nonumber\\
&  =-\underset{i<j}{\sum}\frac{k_{i}\cdot k_{j}}{K}\left[  \ln(z_{j}\cdots
z_{n-2})+\ln\left(  1-z_{i}\cdots z_{j-1}\right)  \right]  ,\text{ }%
K=-k_{1}\cdot k_{2},\\
u\left(  x_{i}\right)   &  =\left(  k^{T}\right)  ^{N-2m-q}\underbrace{\left(
k^{L}\right)  ^{2m}}\left(  k^{\prime L}\right)  ^{q}.(k^{\prime L}%
=\frac{\partial k^{L}}{\partial x_{2}}) \label{uuu}%
\end{align}

In Eq.(\ref{uuu}), we have defined%
\begin{equation}
k=\sum_{i\neq2,n}\frac{k_{i}}{x_{i}-x_{2}}=\sum_{i\neq2,n}\frac{k_{i}}%
{z_{i}\cdots z_{n-2}-z_{2}\cdots z_{n-2}}, \label{48}%
\end{equation}
and $k_{\perp}=\left\vert k_{\perp}\right\vert \sum_{i=1}^{r}e^{T_{i}}%
\omega_{i}=\left\vert k_{\perp}\right\vert e^{\hat{T}}$.

The saddle points $\left(  \tilde{z}_{2},\cdots,\tilde{z}_{n-2}\right)  $ are
the solution of\newline%
\begin{equation}
\frac{\partial f}{\partial z_{2}}=0\text{, }\cdots\text{, }\frac{\partial
f}{\partial z_{n-2}}=0. \label{f22}%
\end{equation}
Note that Eq.(\ref{f22}) implies%
\begin{equation}
\tilde{k}^{L}=\frac{\tilde{k}\cdot k_{2}}{M}=\frac{k_{12}}{M}\left.
\frac{\partial f}{\partial x_{2}}\right\vert _{z_{i}=\tilde{z}_{i}}%
=\frac{k_{12}}{M}\left.  \frac{\partial z_{j}}{\partial x_{2}}\frac{\partial
f}{\partial z_{j}}\right\vert _{z_{i}=\tilde{z}_{i}}=0\text{ , }\left\vert
\tilde{k}\right\vert =\left\vert \tilde{k}_{\perp}\right\vert \text{. }
\label{kkk2}%
\end{equation}
We also define%
\begin{equation}
f_{2}\equiv\frac{\partial f}{\partial z_{2}}\text{, }f_{22}\equiv
\frac{\partial^{2}f}{\partial z_{2}^{2}}\text{, }\tilde{f}=f\left(  \tilde
{z}_{2},\cdots,\tilde{z}_{n-2}\right)  \text{, }\tilde{f}_{22}=\left.
\frac{\partial^{2}f}{\partial z_{2}^{2}}\right\vert _{\left(  \tilde{z}%
_{2},\cdots,\tilde{z}_{n-2}\right)  }.
\end{equation}

In view of the factor $\left(  k^{L}\right)  ^{2m}$ in Eq.(\ref{uuu}) and
Eq.(\ref{kkk2}), all up to $(2m)$-order differentiations of\textbf{ }%
$u$\textbf{ }function in Eq.(\ref{uuu}) at the saddle point vanish except
\cite{NHard}%
\begin{align}
\left.  \frac{\partial^{2m}u}{\partial z_{2}^{2m}}\right\vert _{\left(
\tilde{z}_{2},\cdots,\tilde{z}_{n-2}\right)  }  &  =\left(  \frac{k_{12}}%
{M}\right)  ^{2m+q}\left(  -\sum_{i\neq2,n}\frac{k_{i}^{T}}{\tilde{x}%
_{i}-\tilde{x}_{2}}\right)  ^{N-2m-2q}\left(  2m\right)  !\left(  \tilde
{f}_{22}\right)  ^{q+2m}\nonumber\\
&  =\left(  \frac{k_{12}}{M}\right)  ^{2m+q}\left(  \tilde{k}^{T}\right)
^{N-2m-2q}\left(  2m\right)  !\left(  \tilde{f}_{22}\right)  ^{q+2m}.
\end{align}

Finally, with the saddle point, we can calculate the HSSA to be \cite{NHard}%
\begin{align}
\mathcal{T}^{\left(  N,2m,2q\right)  }  &  =\left(  \prod_{i=3}^{n-2}\int%
_{0}^{1}dz_{i}\text{ }z_{i}^{i-2-N}\right)  \int_{0}^{1}dz_{2}\left(
\frac{\partial^{2m}\tilde{u}}{\partial z_{2}^{2m}}\frac{\left(  z_{2}%
-\tilde{z}_{2}\right)  ^{2m}}{\left(  2m\right)  !}\right)  e^{-Kf}\\
&  \simeq\frac{1}{\left(  2m\right)  !}\frac{\partial^{2m}\tilde{u}}{\partial
z_{2}^{2m}}\left(  \prod_{i=3}^{n-2}\tilde{z}_{i}^{i-2-N}\right)  \int_{0}%
^{1}dz_{2}\left(  z_{2}-\tilde{z}_{2}\right)  ^{2m}e^{-Kf\left(  z_{2}\right)
}\\
&  \simeq\frac{1}{\left(  2m\right)  !}\frac{\partial^{2m}\tilde{u}}{\partial
z_{2}^{2m}}\left(  \prod_{i=3}^{n-2}\tilde{z}_{i}^{i-2-N}\right)  \int%
_{0}^{\infty}dz_{2}\left(  z_{2}-\tilde{z}_{2}\right)  ^{2m}e^{-Kf\left(
z_{2}\right)  }\\
&  =\frac{2\sqrt{\pi}}{m!}\left(  \prod_{i=3}^{n-2}\tilde{z}_{i}%
^{i-2-N}\right)  \frac{e^{-K\tilde{f}}}{\left\vert \tilde{k}\right\vert
^{2m+1}}\left.  \frac{\partial^{2m}u}{\partial z_{2}^{2m}}\right\vert
_{z_{i}=\tilde{z}_{i}}\\
&  =2\sqrt{\pi}e^{-K\tilde{f}}\left\vert \tilde{k}\right\vert ^{N-1}\left(
\prod_{i=3}^{n-2}\tilde{z}_{i}^{i-2-N}\right)  \frac{\left(  2m\right)  !}%
{m!}\left(  \frac{-1}{2M}\right)  ^{2m+q}\left(  \frac{2K\tilde{f}_{22}%
}{\left(  \sum_{i\neq2,n}\frac{k_{i}^{T}}{\tilde{x}_{i}-\tilde{x}_{2}}\right)
^{2}}\right)  ^{m+q}%
\end{align}
where $f\left(  z_{2}\right)  =f\left(  z_{2},\tilde{z}_{3},\cdots,\tilde
{z}_{n-2}\right)  $. The ratios of $n$-point $HSSA$ with $r=1$ is%
\begin{align}
\frac{\mathcal{T}^{\left(  N,m,q\right)  }}{\mathcal{T}^{\left(  N,0,0\right)
}}  &  =\frac{\left(  2m\right)  !}{m!}\left(  \frac{-1}{2M}\right)
^{2m+q}\left(  \frac{2K\tilde{f}_{22}}{\left(  \sum_{i\neq2,n}\frac{k_{i}^{T}%
}{\tilde{x}_{i}-\tilde{x}_{2}}\right)  ^{2}}\right)  ^{m+q}\\
&  =\frac{\left(  2m\right)  !}{m!}\left(  \frac{-1}{2M}\right)  ^{2m+q}%
\end{align}
where the second equality followed from the calculation of decoupling of ZNS
in Eq.(\ref{22}).

This suggests the\textbf{ }identity%
\begin{equation}
\frac{2K\tilde{f}_{22}}{\left(  \sum_{i\neq2,n}\frac{k_{i}^{T}}{\tilde{x}%
_{i}-\tilde{x}_{2}}\right)  ^{2}}=1. \label{aa}%
\end{equation}
For the case of $n=4$, one can easily solve the saddle point $\tilde{z}%
_{2}=\sec^{2}\frac{\phi}{2}$ to verify the identity. We have also proved the
identity for $n=5$ by using maple numerically. Similar proof can be done by
maple for the case of $n=6$.

\subsection{The $n$-point HSSA with $r=2$}

Now we calculate the case of $n$-point HSSA with $r=2$. We want to calculate
$n$-point HSSA with $(n-1)$ tachyons and $1$ high energy state%
\begin{equation}
\left(  \alpha_{-1}^{T_{1}}\right)  ^{N+p_{1}}\left(  \alpha_{-1}^{T_{2}%
}\right)  ^{p_{2}}\left(  \alpha_{-1}^{L}\right)  ^{2m}\left(  \alpha_{-2}%
^{L}\right)  ^{q}\left\vert 0;k\right\rangle \text{, \ }p_{1}+p_{2}=-2(m+q).
\end{equation}
The ratios of $n$-point HSSA with $r=2$ can be similarly calculated to be%
\begin{align}
\frac{\mathcal{T}^{\left(  p_{1},p_{2},m,q\right)  }}{\mathcal{T}^{\left(
N,0,0,0\right)  }}  &  =\frac{\left(  2m\right)  !}{m!}\left(  \frac{-1}%
{2M}\right)  ^{2m+q}\frac{\left(  2K\tilde{f}_{22}\right)  ^{m+q}}{\left(
\sum_{i\neq2,n}\frac{k_{i}^{T_{1}}}{\tilde{x}_{i}-\tilde{x}_{2}}\right)
^{2m+2q+p_{2}}\left(  \sum_{i\neq2,n}\frac{k_{i}^{T_{2}}}{\tilde{x}_{i}%
-\tilde{x}_{2}}\right)  ^{-p_{2}}}\nonumber\\
&  =\frac{\left(  2m\right)  !}{m!}\left(  \frac{-1}{2M}\right)  ^{2m+q}%
\frac{\left(  \frac{\sum_{i\neq2,n}\frac{k_{i}^{T_{2}}}{\tilde{x}_{i}%
-\tilde{x}_{2}}}{\sum_{i\neq2,n}\frac{k_{i}^{T_{1}}}{\tilde{x}_{i}-\tilde
{x}_{2}}}\right)  ^{p_{2}}}{\left(  \frac{\sum_{i\neq2,n}\frac{k_{i}^{T_{1}}%
}{\tilde{x}_{i}-\tilde{x}_{2}}}{\sqrt{2K\tilde{f}_{22}}}\right)  ^{2m+2q}}.
\label{a22}%
\end{align}
On the other hand, the decoupling of ZNS calculated in Eq.(\ref{100}) gives%
\begin{equation}
\frac{\mathcal{T}^{\left(  p_{1},p_{2},m,q\right)  }}{\mathcal{T}^{\left(
N,0,0,0\right)  }}=\frac{\left(  2m\right)  !}{m!}\left(  \frac{-1}%
{2M}\right)  ^{2m+q}\omega_{1}^{p_{1}}\omega_{2}^{p_{2}}=\frac{\left(
2m\right)  !}{m!}\left(  \frac{-1}{2M}\right)  ^{2m+q}\frac{(\tan\theta
_{1})^{p_{2}}}{(\cos\theta_{1})^{2m+2q}}. \label{b22}%
\end{equation}
Eq.(\ref{a22}) and Eq.(\ref{b22}) can be identified for any $p_{2}$, $m$ and
$q$ if%
\begin{equation}
\left(  \sum_{i\neq2,n}\frac{k_{i}^{T_{1}}}{\tilde{x}_{i}-\tilde{x}_{2}%
}\right)  =\sqrt{2K\tilde{f}_{22}}\cos\theta_{1}\text{, }\left(  \sum
_{i\neq2,n}\frac{k_{i}^{T_{2}}}{\tilde{x}_{i}-\tilde{x}_{2}}\right)
=\sqrt{2K\tilde{f}_{22}}\sin\theta_{1},
\end{equation}
which implies the identity%
\begin{equation}
\left(  \sum_{i\neq2,n}\frac{k_{i}^{T_{1}}}{\tilde{x}_{i}-\tilde{x}_{2}%
}\right)  ^{2}+\left(  \sum_{i\neq2,n}\frac{k_{i}^{T_{2}}}{\tilde{x}%
_{i}-\tilde{x}_{2}}\right)  ^{2}=2K\tilde{f}_{22}. \label{qq2}%
\end{equation}
It is not surprising that Eq.(\ref{qq2}) is a generalization of Eq.(\ref{aa})
to two transverse directions $T_{1}$ and $T_{2}$.

\subsection{The $n$-point HSSA with $r\leq24$}

It is now easy to generalize Eq.(\ref{qq2}) to any $r$ (number of $T_{i}$)
with $r\leq24$%
\begin{equation}
\left(  \sum_{i\neq2,n}\frac{k_{i}^{T_{1}}}{\tilde{x}_{i}-\tilde{x}_{2}%
}\right)  ^{2}+\left(  \sum_{i\neq2,n}\frac{k_{i}^{T_{2}}}{\tilde{x}%
_{i}-\tilde{x}_{2}}\right)  ^{2}+\cdots+\left(  \sum_{i\neq2,n}\frac
{k_{i}^{T_{r}}}{\tilde{x}_{i}-\tilde{x}_{2}}\right)  ^{2}=2K\tilde{f}_{22}.
\label{id2}%
\end{equation}
By using Eq.(\ref{48}) and Eq.(\ref{kkk2}), we see that the key identity
Eq.(\ref{id2}) can be written as \cite{NHard}%
\begin{equation}
\tilde{k}^{2}+2M\tilde{k}^{\prime L}=0. \label{iden2}%
\end{equation}
The ratios in Eq.(\ref{100}) are thus proved by the saddle point method.%

%TCIMACRO{\TeXButton{equation number}{\setcounter{equation}{0}
%\renewcommand{\theequation}{\arabic{section}.\arabic{equation}}}}%
%BeginExpansion
\setcounter{equation}{0}
\renewcommand{\theequation}{\arabic{section}.\arabic{equation}}%
%EndExpansion

\section{Stringy scaling of Regge string scattering amplitudes}

Another important high-energy regime of $4$-point SSA is the f\/ixed momentum
transfer regime which contains complementary information of the theory. That
is in the kinematic regime%
\begin{equation}
s\rightarrow\infty,\qquad\sqrt{-t}=\text{f\/ixed,}\quad(\text{but}\ \sqrt
{-t}\neq\infty).
\end{equation}
In this regime, the number of high-energy SSA is much more numerous than that
of the f\/ixed angle regime. One of the reason is that in contrast to the
identification $e^{P}\simeq$ $e^{L}$ in the hard scattering limit, $e^{P}$
\textit{does not} approach to $e^{L}$ in the Regge scattering limit. For
example, at mass level $M^{2}=4$ of open bosonic string, there are only $4$
HSSA while there are $22$ RSSA \cite{bosonic,KLY1}. On the other hand, in the
Regge regime both the saddle-point method and the method of decoupling of
zero-norm states adopted in the calculation of f\/ixed angle regime do not apply.

The complete leading order high-energy open string states in the Regge regime
at each fixed mass level $N=\sum_{n,m,l>0}np_{n}+mq_{m}+lr_{l}$ are%
\begin{equation}
\left\vert v_{n},q_{m},r_{l}\right\rangle =\prod_{n>0}(\alpha_{-n}^{T}%
)^{v_{n}}\prod_{m>0}(\alpha_{-m}^{P})^{q_{m}}\prod_{l>0}(\alpha_{-l}%
^{L})^{r_{l}}|0,k\rangle. \label{RR}%
\end{equation}
It turned out that the $4$-pont RSSA of three tachyons and states in
Eq.(\ref{RR}) are NOT proportional to each other, and the ratios are
$t$-dependent functions. However, it was shown that for the RSSA
$A^{(N,2m,q)}$ with $v_{1}=N-m-q$, $r_{1}=2m$ and $r_{2}=q$ and all others $0$
in Eq.(\ref{RR}), one can extract the ratios of hard string scatterings in
Eq.(\ref{22}) from $A^{(N,2m,q)}$ \cite{RRsusy,LYAM,HLY}. It is thus
reasonable to expect that for the $n$-point ($n\geq5$) RSSA with $n-1$
tachyons and some subset of the high-energy states in Eq.(\ref{HES}), the RSSA
show similar stringy scaling behavior as in Eq.(\ref{100}) of HSSA.

In this paper, we will consider a class of $n$-point ($n\geq5$) RSSA with
$n-1$ tachyons and one high-energy state at mass level $N$
\begin{equation}
\left\vert \left\{  p_{i}\right\}  ,0,0\right\rangle =\left(  \alpha
_{-1}^{T_{1}}\right)  ^{N+p_{1}}\left(  \alpha_{-1}^{T_{2}}\right)  ^{p_{2}%
}\cdots\left(  \alpha_{-1}^{T_{r}}\right)  ^{p_{r}}\left\vert 0;k\right\rangle
, \label{rr}%
\end{equation}
which is obtained by setting $m=q=0$ in Eq.(\ref{HES}). We will show that
these RSSA show stringy scaling behavior for arbitrary $n$ similar to that we
obtained for the HSSA in Eq.(\ref{100}).

There are many different Regge regimes for the $n$-point ($n\geq5$) RSSA. To
specify the Regge regime, we first discuss the system of kinematics variables
we will use. The standard kinematics variables commonly adopted for the
$n$-point scatterings can be defined as following. One first defines the
$(n-3)$ $s$ variables%
\begin{equation}
s_{12}=-\left(  k_{1}+k_{2}\right)  ^{2},s_{123}=-\left(  k_{1}+k_{2}%
+k_{3}\right)  ^{2},\cdots,s_{1,\cdots,n-2}=-\left(  k_{1}+\cdots
+k_{n-2}\right)  ^{2},
\end{equation}
and then defines the $\frac{\left(  n-2\right)  \left(  n-3\right)  }{2}$ $t$
variables
\begin{align}
t_{23}  &  =-\left(  k_{2}+k_{3}\right)  ^{2},t_{24}=-\left(  k_{2}%
+k_{4}\right)  ^{2},\cdots,t_{2,n-1}=-\left(  k_{2}+k_{n-1}\right)
^{2},\nonumber\\
t_{34}  &  =-\left(  k_{3}+k_{4}\right)  ^{2},\cdots,t_{3,n-1}=-\left(
k_{3}+k_{n-1}\right)  ^{2},\nonumber\\
&  \vdots\nonumber\\
t_{n-2,n-1}  &  =-\left(  k_{n-2}+k_{n-1}\right)  ^{2},
\end{align}
which amount to $\frac{n\left(  n-3\right)  }{2}$ independent kinematics variables.

For our purpose in the calculation of this paper, we will adopt another system
of independent kinematics variables. We use the notation $k_{ij}\equiv
k_{i}\cdot k_{j}$ to define the following $\frac{n\left(  n-3\right)  }{2}$
independent kinematics variables%
\begin{align}
&  k_{12},k_{13},k_{14},\cdots,k_{1,n-2},\nonumber\\
&  k_{23},k_{24},k_{25},\cdots,k_{2,n-1},\nonumber\\
&  k_{34},k_{35},\cdots,k_{3,n-1},\nonumber\\
&  \vdots\nonumber\\
&  k_{n-3,n-2},k_{n-3,n-1},\nonumber\\
&  k_{n-2,n-1}. \label{kk}%
\end{align}
For later use, we also define
\begin{equation}
k_{1,\cdots,i-1,i}=k_{1,\cdots,i-1}+\sum_{j=1}^{i-1}k_{ji}, \label{kk2}%
\end{equation}
which means, for example,%
\begin{equation}
k_{123}=k_{12}+k_{13}+k_{23},k_{1234}=k_{123}+k_{14}+k_{24}+k_{34}%
,k_{12345}=k_{1234}+k_{15}+k_{25}+k_{35}+k_{45}.
\end{equation}

\subsection{The $5$-point and $6$-point Regge stringy scaling}

Let's begin with the calculation of $5$-point RSSA with $r=2$ in
Eq.(\ref{rr}). The kinematics are%
\begin{align}
k_{1}  &  =\left(  \sqrt{p^{2}+M_{1}^{2}},-p,0,0\right)  ,\nonumber\\
k_{2}  &  =\left(  \sqrt{p^{2}+M_{2}^{2}},p,0,0\right)  ,\nonumber\\
k_{3}  &  =\left(  -\sqrt{q_{3}^{2}+M_{3}^{2}},-q_{3}\cos\phi_{1}^{3}%
,-q_{3}\sin\phi_{1}^{3},0\right)  ,\nonumber\\
k_{4}  &  =\left(  -\sqrt{q_{4}^{2}+M_{4}^{2}},-q_{4}\cos\phi_{1}^{4}%
,-q_{4}\sin\phi_{1}^{4}\cos\phi_{2}^{4},-q_{4}\sin\phi_{1}^{4}\sin\phi_{2}%
^{4}\right)  ,\nonumber\\
k_{5}  &  =\left(  -\sqrt{q_{5}^{2}+M_{5}^{2}},-q_{5}\cos\phi_{1}^{5}%
,-q_{5}\sin\phi_{1}^{5}\cos\phi_{2}^{5},-q_{5}\sin\phi_{1}^{5}\sin\phi_{2}%
^{5}\right)  .
\end{align}
During the calculation, we will keep record of the notations used for each
step so that eventually we can generalize the calculation to the case of
$n$-point RSSA. The amplitude of state
\begin{equation}
\left(  \alpha_{-1}^{T_{1}}\right)  ^{N+p_{1}}\left(  \alpha_{-1}^{T_{2}%
}\right)  ^{p_{2}}\left\vert 0,k\right\rangle ,p_{1}+p_{2}=0
\end{equation}
and $4$ tachyon states can be written as%
\begin{align}
A^{\left\{  p_{1},p_{2}\right\}  ,0,0}  &  =\int_{0}^{1}dx_{3}\int_{0}^{x_{3}%
}dx_{2}\times x_{2}^{k_{12}}x_{3}^{k_{13}}\left(  x_{3}-x_{2}\right)
^{k_{23}}\left(  1-x_{2}\right)  ^{k_{24}}\left(  1-x_{3}\right)  ^{k_{34}%
}\nonumber\\
&  \times\left[  \frac{k_{3}^{T_{1}}}{x_{3}-x_{2}}+\frac{k_{4}^{T_{1}}%
}{1-x_{2}}\right]  ^{N+p_{1}}\left[  \underset{0}{\underbrace{\frac
{k_{3}^{T_{2}}}{x_{3}-x_{2}}}}+\frac{k_{4}^{T_{2}}}{1-x_{2}}\right]  ^{p_{2}}.
\end{align}
One can easily find that $k_{3}^{T_{2}}=0$. After doing the change of
variables
\begin{equation}
x_{2}=z_{2}z_{3},x_{3}=z_{3},
\end{equation}
we can rewrite the above $5$-point amplitude as following%
\begin{align}
A^{\left\{  p_{1},p_{2}\right\}  ,0,0}  &  =\int_{0}^{1}dz_{3}\int_{0}%
^{1}dz_{2}z_{2}^{k_{12}}z_{3}^{k_{123}+1}\left(  1-z_{2}\right)  ^{k_{23}%
}\left(  1-z_{2}z_{3}\right)  ^{k_{24}}\left(  1-z_{3}\right)  ^{k_{34}%
}\nonumber\\
&  \times\left[  \frac{k_{3}^{T_{1}}}{z_{3}-z_{2}z_{3}}+\frac{k_{4}^{T_{1}}%
}{1-z_{2}z_{3}}\right]  ^{N+p_{1}}\left[  \frac{k_{4}^{T_{2}}}{1-z_{2}z_{3}%
}\right]  ^{p_{2}}%
\end{align}
where we have defined $k_{123}=k_{12}+k_{23}+k_{13}$. Next, let's perform the
binomial expansion on the bracket to obtain
\begin{align}
&  A^{\left\{  p_{1},p_{2}\right\}  ,0,0}=\sum_{J_{1}^{1}+J_{2}^{1}=N+p_{1}%
}\frac{\left(  N+p_{1}\right)  !}{J_{1}^{1}!J_{2}^{1}!}\left(  k_{3}^{T_{1}%
}\right)  ^{J_{1}^{1}}\left(  k_{4}^{T_{1}}\right)  ^{J_{2}^{1}}\left(
k_{4}^{T_{2}}\right)  ^{p_{2}}\nonumber\\
&  \times\int_{0}^{1}dz_{3}\int_{0}^{1}dz_{2}z_{2}^{k_{12}}z_{3}%
^{k_{123}+1-J_{1}^{1}}\left(  1-z_{2}\right)  ^{k_{23}-J_{1}^{1}}\left(
1-z_{2}z_{3}\right)  ^{k_{24}-J_{2}^{1}-p_{2}}\left(  1-z_{3}\right)
^{k_{34}}.
\end{align}
For the next step, we expand the crossing term $\left(  1-z_{2}z_{3}\right)
^{k_{24}-J_{2}^{1}-p_{2}}$ to obtain%
\begin{align}
&  A^{\left\{  p_{1},p_{2}\right\}  ,0,0}=\sum_{J_{1}^{1}+J_{2}^{1}=N+p_{1}%
}\frac{\left(  N+p_{1}\right)  !}{J_{1}^{1}!J_{2}^{1}!}\left(  k_{3}^{T_{1}%
}\right)  ^{J_{1}^{1}}\left(  k_{4}^{T_{1}}\right)  ^{J_{2}^{1}}\left(
k_{4}^{T_{2}}\right)  ^{p_{2}}\nonumber\\
&  \times\sum_{m_{23}}\frac{\left(  -k_{24}+p_{2}+J_{2}^{1}\right)  _{m_{23}}%
}{m_{23}!}\int_{0}^{1}dz_{2}z_{2}^{k_{12}+m_{23}}\left(  1-z_{2}\right)
^{k_{23}-J_{1}^{1}}\int_{0}^{1}dz_{3}z_{3}^{k_{123}+1-J_{1}^{1}+m_{23}}\left(
1-z_{3}\right)  ^{k_{34}}%
\end{align}
where the subscripts of $m_{23}$ keep record of the subscripts $z_{2}z_{3}$ in
$\left(  1-z_{2}z_{3}\right)  ^{k_{24}-J_{2}^{1}-p_{2}}$. After the
integration, the amplitude can be written as
\begin{align}
A^{\left\{  p_{1},p_{2}\right\}  ,0,0}  &  =\sum_{J_{1}^{1}+J_{2}^{1}=N+p_{1}%
}\frac{\left(  N+p_{1}\right)  !}{J_{1}^{1}!J_{2}^{1}!}\left(  k_{3}^{T_{1}%
}\right)  ^{J_{1}^{1}}\left(  k_{4}^{T_{1}}\right)  ^{J_{2}^{1}}\left(
k_{4}^{T_{2}}\right)  ^{p_{2}}\nonumber\\
&  \times\sum_{m_{23}}\frac{\left(  -k_{24}+p_{2}+J_{2}^{1}\right)  _{m_{23}}%
}{m_{23}!}\frac{\Gamma\left(  k_{12}+1+m_{23}\right)  \Gamma\left(
k_{23}+1-J_{1}^{1}\right)  }{\Gamma\left(  k_{12}+k_{23}+2+m_{23}-J_{1}%
^{1}\right)  }\nonumber\\
&  \times\frac{\Gamma\left(  k_{123}+2+m_{23}-J_{1}^{1}\right)  \Gamma\left(
k_{34}+1\right)  }{\Gamma\left(  k_{123}+k_{34}+3+m_{23}-J_{1}^{1}\right)  }.
\end{align}

Now we choose to work on the Regge regime defined by%
\begin{equation}
k_{123}\sim s,k_{34}\sim s,k_{123}+k_{34}\sim t
\end{equation}
where $s\rightarrow\infty$ and $t=$ fixed. (we will use these notations to
define a Regge regime for the rest of the paper) In this Regge regime, the
amplitude can be approximated as%
\begin{align}
A^{\left\{  p_{1},p_{2}\right\}  ,0,0}  &  \sim\sum_{J_{1}^{1}+J_{2}%
^{1}=N+p_{1}}\frac{\left(  N+p_{1}\right)  !}{J_{1}^{1}!J_{2}^{1}!}\left(
k_{3}^{T_{1}}\right)  ^{J_{1}^{1}}\left(  k_{4}^{T_{1}}\right)  ^{J_{2}^{1}%
}\left(  k_{4}^{T_{2}}\right)  ^{p_{2}}\nonumber\\
&  \times\sum_{m_{23}}\frac{\left(  -k_{24}+p_{2}+J_{2}^{1}\right)  _{m_{23}}%
}{m_{23}!}\frac{\Gamma\left(  k_{12}+1+m_{23}\right)  \Gamma\left(
k_{23}+1-J_{1}^{1}\right)  }{\Gamma\left(  k_{12}+k_{23}+2+m_{23}-J_{1}%
^{1}\right)  }\nonumber\\
&  \times\frac{\left(  k_{123}\right)  ^{m_{23}-J_{1}^{1}}\Gamma\left(
k_{123}+2\right)  \Gamma\left(  k_{34}+1\right)  }{\left(  k_{123}%
+k_{34}+3\right)  _{m_{23}-J_{1}^{1}}\Gamma\left(  k_{123}+k_{34}+3\right)  }.
\end{align}
The leading power of $k_{123}$ occurs when $J_{1}^{1}=0$ which means
$J_{2}^{1}=N+p_{1}$. Since $p_{1}+p_{2}=0$, the leading term of the RSSA is
\begin{align}
A^{\left\{  p_{1},p_{2}\right\}  ,0,0}  &  \sim\left(  k_{4}^{T_{1}}\right)
^{N+p_{1}}\left(  k_{4}^{T_{2}}\right)  ^{p_{2}}\sum_{m_{23}}\frac{\left(
-k_{24}+N\right)  _{m_{23}}}{m_{23}!}\frac{\Gamma\left(  k_{12}+1+m_{23}%
\right)  \Gamma\left(  k_{23}+1\right)  }{\Gamma\left(  k_{12}+k_{23}%
+2+m_{23}\right)  }\nonumber\\
&  \times\frac{\left(  k_{123}\right)  ^{m_{23}}\Gamma\left(  k_{123}%
+2\right)  \Gamma\left(  k_{34}+1\right)  }{\left(  k_{123}+k_{34}+3\right)
_{m_{23}}\Gamma\left(  k_{123}+k_{34}+3\right)  }.
\end{align}
The ratio of $A^{\left\{  p_{1},p_{2}\right\}  ,0,0}$ and $A^{\left\{
0,0\right\}  ,0,0}$ can be easily calculated to be
\begin{align}
\frac{A^{\left\{  p_{1},p_{2}\right\}  ,0,0}}{A^{\left\{  0,0\right\}  ,0,0}}
&  =\frac{\left(  k_{4}^{T_{1}}\right)  ^{N+p_{1}}\left(  k_{4}^{T_{2}%
}\right)  ^{p_{2}}}{\left(  k_{4}^{T_{1}}\right)  ^{N}}=\left(  k_{4}^{T_{1}%
}\right)  ^{p_{1}}\left(  k_{4}^{T_{2}}\right)  ^{p_{2}}=\left(  -q_{4}%
\sin\phi_{1}^{4}\cos\phi_{2}^{4}\right)  ^{p_{1}}\left(  -q_{4}\sin\phi
_{1}^{4}\sin\phi_{2}^{4}\right)  ^{p_{2}}\nonumber\\
&  =\left(  \cos\theta_{1}\right)  ^{p_{1}}\left(  \sin\theta_{1}\right)
^{p_{2}}=\left(  \omega_{1}\right)  ^{p_{1}}\left(  \omega_{2}\right)
^{p_{2}},
\end{align}
which is the same as Eq.(\ref{100}) with $m=q=0$ and $r=2$.

Let's now calculate the $6$-point RSSA with $r=3$ in Eq.(\ref{rr}). The
kinematics are%
\begin{align}
k_{1}  &  =\left(  \sqrt{p^{2}+M_{1}^{2}},-p,0,0,0\right)  ,\nonumber\\
k_{2}  &  =\left(  \sqrt{p^{2}+M_{2}^{2}},p,0,0,0\right)  ,\nonumber\\
k_{3}  &  =\left(  -\sqrt{q_{3}^{2}+M_{3}^{2}},-q_{3}\cos\phi_{1}^{3}%
,-q_{3}\sin\phi_{1}^{3},0,0\right)  ,\nonumber\\
k_{4}  &  =\left(  -\sqrt{q_{4}^{2}+M_{4}^{2}},-q_{4}\cos\phi_{1}^{4}%
,-q_{4}\sin\phi_{1}^{4}\cos\phi_{2}^{4},-q_{4}\sin\phi_{1}^{4}\sin\phi_{2}%
^{4},0\right)  ,\nonumber\\
k_{5}  &  =\left(  -\sqrt{q_{5}^{2}+M_{5}^{2}},-q_{5}\cos\phi_{1}^{5}%
,-q_{5}\sin\phi_{1}^{5}\cos\phi_{2}^{5},-q_{5}\sin\phi_{1}^{5}\sin\phi_{2}%
^{5}\cos\phi_{3}^{5},-q_{5}\sin\phi_{1}^{5}\sin\phi_{2}^{5}\sin\phi_{3}%
^{5}\right)  ,\nonumber\\
k_{6}  &  =\left(  -\sqrt{q_{5}^{2}+M_{5}^{2}},-q_{6}\cos\phi_{1}^{6}%
,-q_{6}\sin\phi_{1}^{6}\cos\phi_{2}^{6},-q_{6}\sin\phi_{1}^{6}\sin\phi_{2}%
^{6}\cos\phi_{3}^{6},-q_{6}\sin\phi_{1}^{6}\sin\phi_{2}^{6}\sin\phi_{3}%
^{6}\right)  .
\end{align}
The amplitude of state
\begin{equation}
\left(  \alpha_{-1}^{T_{1}}\right)  ^{N+p_{1}}\left(  \alpha_{-1}^{T_{2}%
}\right)  ^{p_{2}}\left(  \alpha_{-1}^{T_{3}}\right)  ^{p_{3}}\left\vert
0,k\right\rangle ,p_{1}+p_{2}+p_{3}=0
\end{equation}
and $5$ tachyon states is%

\begin{align}
&  A^{\left\{  p_{1},p_{2},p_{3}\right\}  ,0,0}\nonumber\\
&  =\int_{0}^{1}dx_{4}\int_{0}^{x_{4}}dx_{3}\int_{0}^{x_{3}}dx_{2}\text{
}x_{2}^{k_{12}}x_{3}^{k_{13}}x_{4}^{k_{14}}\left(  x_{3}-x_{2}\right)
^{k_{23}}\left(  x_{4}-x_{2}\right)  ^{k_{24}}\left(  1-x_{2}\right)
^{k_{25}}\left(  x_{4}-x_{3}\right)  ^{k_{34}}\left(  1-x_{3}\right)
^{k_{35}}\left(  1-x_{4}\right)  ^{k_{45}}\nonumber\\
&  \times\left[  \frac{k_{3}^{T_{1}}}{x_{3}-x_{2}}+\frac{k_{4}^{T_{1}}}%
{x_{4}-x_{2}}+\frac{k_{5}^{T_{1}}}{\underset{x_{5}}{\underbrace{1}}-x_{2}%
}\right]  ^{N+p_{1}}\left[  \underset{=0}{\underbrace{\frac{k_{3}^{T_{2}}%
}{x_{3}-x_{2}}}}+\frac{k_{4}^{T_{2}}}{x_{4}-x_{2}}+\frac{k_{5}^{T_{2}}%
}{1-x_{2}}\right]  ^{p_{2}}\left[  \underset{=0\text{ }}{\underbrace{\frac
{k_{3}^{T_{3}}}{x_{3}-x_{2}}}}+\underset{=0\text{ }}{\underbrace{\frac
{k_{4}^{T_{3}}}{x_{4}-x_{2}}}}+\frac{k_{5}^{T_{3}}}{1-x_{2}}\right]  ^{p_{3}}.
\end{align}
Since $k_{3}^{T_{2}}=k_{3}^{T_{3}}=k_{4}^{T_{3}}=0,$ we can rewrite the
amplitude as%
\begin{align}
A^{\left\{  p_{1},p_{2},p_{3}\right\}  ,0,0}  &  =\int_{0}^{1}dx_{4}\int%
_{0}^{x_{4}}dx_{3}\int_{0}^{x_{3}}dx_{2}\text{ }x_{2}^{k_{12}}x_{3}^{k_{13}%
}x_{4}^{k_{14}}\left(  x_{3}-x_{2}\right)  ^{k_{23}}\left(  x_{4}%
-x_{2}\right)  ^{k_{24}}\left(  1-x_{2}\right)  ^{k_{25}}\left(  x_{4}%
-x_{3}\right)  ^{k_{34}}\left(  1-x_{3}\right)  ^{k_{35}}\left(
1-x_{4}\right)  ^{k_{45}}\nonumber\\
&  \times\left[  \frac{k_{3}^{T_{1}}}{x_{3}-x_{2}}+\frac{k_{4}^{T_{1}}}%
{x_{4}-x_{2}}+\frac{k_{5}^{T_{1}}}{1-x_{2}}\right]  ^{N+p_{1}}\left[
\frac{k_{4}^{T_{2}}}{x_{4}-x_{2}}+\frac{k_{5}^{T_{2}}}{1-x_{2}}\right]
^{p_{2}}\left[  \frac{k_{5}^{T_{3}}}{1-x_{2}}\right]  ^{p_{3}}.
\end{align}
We can do the following change of variables
\begin{equation}
x_{i}=z_{i}\cdots z_{n-2},
\end{equation}
or%
\begin{equation}
x_{2}=z_{2}z_{3}z_{4},x_{3}=z_{3}z_{4},x_{4}=z_{4}%
\end{equation}
to obtain%
\begin{align}
A^{\left\{  p_{1},p_{2},p_{3}\right\}  ,0,0}  &  =\int_{0}^{1}dz_{4}\int%
_{0}^{1}dz_{3}\int_{0}^{1}dz_{2}\times z_{2}^{k_{12}}z_{3}^{k_{123}+1}%
z_{4}^{k_{1234}+2}\left(  1-z_{2}\right)  ^{k_{23}}\left(  1-z_{3}\right)
^{k_{34}}\left(  1-z_{4}\right)  ^{k_{45}}\nonumber\\
&  \times\left(  1-z_{2}z_{3}\right)  ^{k_{24}}\left(  1-z_{2}z_{3}%
z_{4}\right)  ^{k_{25}}\left(  1-z_{3}z_{4}\right)  ^{k_{35}}\nonumber\\
&  \times\left[  \frac{k_{3}^{T_{1}}}{z_{3}z_{4}-z_{2}z_{3}z_{4}}+\frac
{k_{4}^{T_{1}}}{z_{4}-z_{2}z_{3}z_{4}}+\frac{k_{5}^{T_{1}}}{1-z_{2}z_{3}z_{4}%
}\right]  ^{N+p_{1}}\left[  \frac{k_{4}^{T_{2}}}{z_{4}-z_{2}z_{3}z_{4}}%
+\frac{k_{5}^{T_{2}}}{1-z_{2}z_{3}z_{4}}\right]  ^{p_{2}}\left[  \frac
{k_{5}^{T_{3}}}{1-z_{2}z_{3}z_{4}}\right]  ^{p_{3}}%
\end{align}
where we have defined
\begin{equation}
k_{123}=k_{12}+k_{13}+k_{23},k_{1234}=k_{12}+k_{13}+k_{14}+k_{23}%
+k_{24}+k_{34}.
\end{equation}

Next, let's perform the binomial expansion on the brackets to obtain%
\begin{align}
A^{\left\{  p_{1},p_{2},p_{3}\right\}  ,0,0}  &  =\int_{0}^{1}dz_{4}\int%
_{0}^{1}dz_{3}\int_{0}^{1}dz_{2}\times z_{2}^{k_{12}}z_{3}^{k_{123}+1}%
z_{4}^{k_{1234}+2}\left(  1-z_{2}\right)  ^{k_{23}}\left(  1-z_{3}\right)
^{k_{34}}\left(  1-z_{4}\right)  ^{k_{45}}\nonumber\\
&  \times\left(  1-z_{2}z_{3}\right)  ^{k_{24}}\left(  1-z_{2}z_{3}%
z_{4}\right)  ^{k_{25}}\left(  1-z_{3}z_{4}\right)  ^{k_{35}}\nonumber\\
&  \times\sum_{J_{1}^{1}+J_{2}^{1}+J_{3}^{1}=N+p_{1}}^{N+p_{1}}\frac{\left(
N+p_{1}\right)  !}{J_{1}^{1}!J_{2}^{1}!J_{3}^{1}!}\left(  \frac{k_{3}^{T_{1}}%
}{z_{3}z_{4}-z_{2}z_{3}z_{4}}\right)  ^{J_{1}^{1}}\left(  \frac{k_{4}^{T_{1}}%
}{z_{4}-z_{2}z_{3}z_{4}}\right)  ^{J_{2}^{1}}\left(  \frac{k_{5}^{T_{1}}%
}{1-z_{2}z_{3}z_{4}}\right)  ^{J_{3}^{1}}\nonumber\\
&  \times\sum_{J_{1}^{2}+J_{2}^{2}=p_{2}}^{p_{2}}\frac{p_{2}!}{J_{1}^{2}%
!J_{2}^{2}!}\left(  \frac{k_{4}^{T_{2}}}{z_{4}-z_{2}z_{3}z_{4}}\right)
^{J_{1}^{2}}\left(  \frac{k_{5}^{T_{2}}}{1-z_{2}z_{3}z_{4}}\right)
^{J_{2}^{2}}\left[  \frac{k_{5}^{T_{3}}}{1-z_{2}z_{3}z_{4}}\right]  ^{p_{3}}%
\end{align}
where $J_{1}^{1}$ ,$J_{2}^{1}$ ,$J_{3}^{1}$,$J_{1}^{2}$,$J_{2}^{2}$ are
non-negative integers with $J_{1}^{1}+J_{2}^{1}=N+p_{1}$ and $J_{1}^{2}%
+J_{2}^{2}=p_{2}$. We then rearrange the above equation
\begin{align}
A^{\left\{  p_{1},p_{2},p_{3}\right\}  ,0,0}  &  =\sum_{J_{1}^{1}+J_{2}%
^{1}+J_{3}^{1}=N+p_{1}}^{N+p_{1}}\frac{\left(  N+p_{1}\right)  !}{J_{1}%
^{1}!J_{2}^{1}!J_{3}^{1}!}\left(  k_{3}^{T_{1}}\right)  ^{J_{1}^{1}}\left(
k_{4}^{T_{1}}\right)  ^{J_{2}^{1}}\left(  k_{5}^{T_{1}}\right)  ^{J_{3}^{1}%
}\sum_{J_{1}^{2}+J_{2}^{2}=p_{2}}^{p_{2}}\frac{p_{2}!}{J_{1}^{2}!J_{2}^{2}%
!}\left(  k_{4}^{T_{2}}\right)  ^{J_{1}^{2}}\left(  k_{5}^{T_{2}}\right)
^{J_{2}^{2}}\left(  k_{5}^{T_{3}}\right)  ^{p_{3}}\nonumber\\
&  \times\int_{0}^{1}dz_{4}\int_{0}^{1}dz_{3}\int_{0}^{1}dz_{2}\times
z_{2}^{k_{12}}z_{3}^{k_{123}+1-J_{1}^{1}}z_{4}^{k_{1234}+2-J_{1}^{1}-\left(
J_{2}^{1}+J_{1}^{2}\right)  }\left(  1-z_{2}\right)  ^{k_{23}-J_{1}^{1}%
}\left(  1-z_{3}\right)  ^{k_{34}}\left(  1-z_{4}\right)  ^{k_{45}}\nonumber\\
&  \times\left(  1-z_{2}z_{3}\right)  ^{k_{24}-\left(  J_{2}^{1}+J_{1}%
^{2}\right)  }\left(  1-z_{2}z_{3}z_{4}\right)  ^{k_{25}-\left(  J_{3}%
^{1}+J_{2}^{2}+p_{3}\right)  }\left(  1-z_{3}z_{4}\right)  ^{k_{35}},
\end{align}
and expand the crossing terms to obain%
\begin{align}
A^{\left\{  p_{1},p_{2},p_{3}\right\}  ,0,0}  &  =\sum_{J_{1}^{1}+J_{2}%
^{1}+J_{3}^{1}=N+p_{1}}^{N+p_{1}}\frac{\left(  N+p_{1}\right)  !}{J_{1}%
^{1}!J_{2}^{1}!J_{3}^{1}!}\left(  k_{3}^{T_{1}}\right)  ^{J_{1}^{1}}\left(
k_{4}^{T_{1}}\right)  ^{J_{2}^{1}}\left(  k_{5}^{T_{1}}\right)  ^{J_{3}^{1}%
}\sum_{J_{1}^{2}+J_{2}^{2}=p_{2}}^{p_{2}}\frac{p_{2}!}{J_{1}^{2}!J_{2}^{2}%
!}\left(  k_{4}^{T_{2}}\right)  ^{J_{1}^{2}}\left(  k_{5}^{T_{2}}\right)
^{J_{2}^{2}}\left(  k_{5}^{T_{3}}\right)  ^{p_{3}}\nonumber\\
&  \times\int_{0}^{1}dz_{4}\int_{0}^{1}dz_{3}\int_{0}^{1}dz_{2}\times
z_{2}^{k_{12}}z_{3}^{k_{123}+1-J_{1}^{1}}z_{4}^{k_{1234}+2-J_{1}^{1}-\left(
J_{2}^{1}+J_{1}^{2}\right)  }\left(  1-z_{2}\right)  ^{k_{23}-J_{1}^{1}%
}\left(  1-z_{3}\right)  ^{k_{34}}\left(  1-z_{4}\right)  ^{k_{45}}\nonumber\\
&  \times\sum_{m_{23}=0}\frac{\left[  -k_{24}+\left(  J_{2}^{1}+J_{1}%
^{2}\right)  \right]  _{m_{23}}}{m_{23}!}\left(  z_{2}z_{3}\right)  ^{m_{23}%
}\sum_{m_{24}=0}\frac{\left[  -k_{25}+\left(  J_{3}^{1}+J_{2}^{2}%
+p_{3}\right)  \right]  _{m_{24}}}{m_{24}!}\left(  z_{2}z_{3}z_{4}\right)
^{m_{24}}\nonumber\\
&  \times\sum_{m_{34}=0}\frac{\left[  -k_{35}\right]  _{m_{34}}}{m_{34}%
!}\left(  z_{3}z_{4}\right)  ^{m_{34}}%
\end{align}
where, for example, the subscripts of $m_{24}$ keep record of the first and
the last subscripts of $\left(  z_{2}z_{3}z_{4}\right)  $ etc.. We rearrange
the above equation again%
\begin{align}
A^{\left\{  p_{1},p_{2},p_{3}\right\}  ,0,0}  &  =\sum_{J_{1}^{1}+J_{2}%
^{1}+J_{3}^{1}=N+p_{1}}^{N+p_{1}}\frac{\left(  N+p_{1}\right)  !}{J_{1}%
^{1}!J_{2}^{1}!J_{3}^{1}!}\left(  k_{3}^{T_{1}}\right)  ^{J_{1}^{1}}\left(
k_{4}^{T_{1}}\right)  ^{J_{2}^{1}}\left(  k_{5}^{T_{1}}\right)  ^{J_{3}^{1}%
}\sum_{J_{1}^{2}+J_{2}^{2}=p_{2}}^{N+p_{1}}\left(  k_{4}^{T_{2}}\right)
^{J_{1}^{2}}\left(  k_{5}^{T_{2}}\right)  ^{J_{2}^{2}}\left(  k_{5}^{T_{3}%
}\right)  ^{p_{3}}\nonumber\\
&  \times\sum_{m_{23}=0}\frac{\left[  -k_{24}+\left(  J_{2}^{1}+J_{1}%
^{2}\right)  \right]  _{m_{23}}}{m_{23}!}\sum_{m_{24}=0}\frac{\left[
-k_{25}+\left(  J_{3}^{1}+J_{2}^{2}+p_{3}\right)  \right]  _{m_{24}}}{m_{24}%
!}\sum_{m_{34}=0}\frac{\left[  -k_{35}\right]  _{m_{34}}}{m_{34}!}\nonumber\\
&  \times\int_{0}^{1}dz_{2}z_{2}^{k_{12}+m_{23}+m_{24}}\left(  1-z_{2}\right)
^{k_{23}-J_{1}^{1}}\nonumber\\
&  \times\int_{0}^{1}dz_{3}z_{3}^{k_{123}+1-J_{1}^{1}+m_{23}+m_{24}+m_{34}%
}\left(  1-z_{3}\right)  ^{k_{34}}\nonumber\\
&  \times\int_{0}^{1}dz_{4}z_{4}^{k_{1234}+2-J_{1}^{1}-\left(  J_{2}^{1}%
+J_{1}^{2}\right)  +m_{24}+m_{34}}\left(  1-z_{4}\right)  ^{k_{45}},
\end{align}
and perform the integration to obtain
\begin{align}
A^{\left\{  p_{1},p_{2},p_{3}\right\}  ,0,0}  &  =\sum_{J_{1}^{1}+J_{2}%
^{1}+J_{3}^{1}=N+p_{1}}^{N+p_{1}}\frac{\left(  N+p_{1}\right)  !}{J_{1}%
^{1}!J_{2}^{1}!J_{3}^{1}!}\left(  k_{3}^{T_{1}}\right)  ^{J_{1}^{1}}\left(
k_{4}^{T_{1}}\right)  ^{J_{2}^{1}}\left(  k_{5}^{T_{1}}\right)  ^{J_{3}^{1}%
}\sum_{J_{1}^{2}+J_{2}^{2}=p_{2}}^{p_{2}}\frac{p_{2}!}{J_{1}^{2}!J_{2}^{2}%
!}\left(  k_{4}^{T_{2}}\right)  ^{J_{1}^{2}}\left(  k_{5}^{T_{2}}\right)
^{J_{2}^{2}}\left(  k_{5}^{T_{3}}\right)  ^{p_{3}}\nonumber\\
&  \times\sum_{m_{23}=0}\frac{\left[  -k_{24}+\left(  J_{2}^{1}+J_{1}%
^{2}\right)  \right]  _{m_{23}}}{m_{23}!}\sum_{m_{24}=0}\frac{\left[
-k_{25}+\left(  J_{3}^{1}+J_{2}^{2}+p_{3}\right)  \right]  _{m_{24}}}{m_{24}%
!}\sum_{m_{34}=0}\frac{\left[  -k_{35}\right]  _{m_{34}}}{m_{34}!}\nonumber\\
&  \times\frac{\Gamma\left(  k_{12}+1+m_{23}+m_{24}\right)  \Gamma\left(
k_{23}+1-J_{1}^{1}\right)  }{\Gamma\left(  k_{12}+k_{23}+2-J_{1}^{1}%
+m_{23}+m_{24}\right)  }\nonumber\\
&  \times\frac{\Gamma\left(  k_{123}+2-J_{1}^{1}+m_{23}+m_{24}+m_{34}\right)
\Gamma\left(  k_{34}+1\right)  }{\Gamma\left(  k_{123}+k_{34}+3-J_{1}%
^{1}+m_{23}+m_{24}+m_{34}\right)  }\nonumber\\
&  \times\frac{\Gamma\left(  k_{1234}+3-\left(  J_{1}^{1}+J_{2}^{1}+J_{1}%
^{2}\right)  +m_{24}+m_{34}\right)  \Gamma\left(  k_{45}+1\right)  }%
{\Gamma\left(  k_{1234}+k_{23}+4-\left(  J_{1}^{1}+J_{2}^{1}+J_{1}^{2}\right)
+m_{24}+m_{34}\right)  }.
\end{align}

Now we choose to work on the Regge regime defined by%

\begin{equation}
k_{1234}\sim s,k_{1234}+k_{23}\sim t.
\end{equation}
In this Regge regime, the amplitude can be approximated as%
\begin{align}
A^{\left\{  p_{1},p_{2},p_{3}\right\}  ,0,0}  &  \sim\sum_{J_{1}^{1}+J_{2}%
^{1}+J_{3}^{1}=N+p_{1}}^{N+p_{1}}\frac{\left(  N+p_{1}\right)  !}{J_{1}%
^{1}!J_{2}^{1}!J_{3}^{1}!}\left(  k_{3}^{T_{1}}\right)  ^{J_{1}^{1}}\left(
k_{4}^{T_{1}}\right)  ^{J_{2}^{1}}\left(  k_{5}^{T_{1}}\right)  ^{J_{3}^{1}%
}\sum_{J_{1}^{2}+J_{2}^{2}=p_{2}}^{p_{2}}\frac{p_{2}!}{J_{1}^{2}!J_{2}^{2}%
!}\left(  k_{4}^{T_{2}}\right)  ^{J_{1}^{2}}\left(  k_{5}^{T_{2}}\right)
^{J_{2}^{2}}\left(  k_{5}^{T_{3}}\right)  ^{p_{3}}\nonumber\\
&  \times\sum_{m_{23}=0}\frac{\left[  -k_{24}+\left(  J_{2}^{1}+J_{1}%
^{2}\right)  \right]  _{m_{23}}}{m_{23}!}\sum_{m_{24}=0}\frac{\left[
-k_{25}+\left(  J_{3}^{1}+J_{2}^{2}+p_{3}\right)  \right]  _{m_{24}}}{m_{24}%
!}\sum_{m_{34}=0}\frac{\left[  -k_{35}\right]  _{m_{34}}}{m_{34}!}\nonumber\\
&  \times\frac{\Gamma\left(  k_{12}+1+m_{23}+m_{24}\right)  \Gamma\left(
k_{23}+1-J_{1}^{1}\right)  }{\Gamma\left(  k_{12}+k_{23}+2-J_{1}^{1}%
+m_{23}+m_{24}\right)  }\nonumber\\
&  \times\frac{\Gamma\left(  k_{123}+2-J_{1}^{1}+m_{23}+m_{24}+m_{34}\right)
\Gamma\left(  k_{34}+1\right)  }{\Gamma\left(  k_{123}+k_{34}+3-J_{1}%
^{1}+m_{23}+m_{24}+m_{34}\right)  }\nonumber\\
&  \times\frac{\left(  k_{1234}\right)  ^{-\left(  J_{1}^{1}+J_{2}^{1}%
+J_{1}^{2}\right)  +m_{24}+m_{34}}}{\left(  k_{1234}+k_{23}+4\right)
_{-\left(  J_{1}^{1}+J_{2}^{1}+J_{1}^{2}\right)  +m_{24}+m_{34}}}\frac
{\Gamma\left(  k_{1234}+3\right)  \Gamma\left(  k_{45}+1\right)  }%
{\Gamma\left(  k_{1234}+k_{23}+4\right)  }.
\end{align}
We can now take $J_{1}^{1}=J_{2}^{1}=J_{1}^{2}=0$ to extract the leading order
term in $k_{1234}$. This inplies $J_{3}^{1}=N+p_{1}$ and $J_{2}^{2}=p_{2}$
which give%
\begin{align}
A^{\left\{  p_{1},p_{2},p_{3}\right\}  ,0,0}  &  \sim\left(  k_{5}^{T_{1}%
}\right)  ^{N+p_{1}}\left(  k_{5}^{T_{2}}\right)  ^{p_{2}}\left(  k_{5}%
^{T_{3}}\right)  ^{p_{3}}\nonumber\\
&  \times\sum_{m_{23}=0}\frac{\left[  -k_{24}\right]  _{m_{23}}}{m_{23}!}%
\sum_{m_{24}=0}\frac{\left[  -k_{25}+N\right]  _{m_{24}}}{m_{24}!}\sum
_{m_{34}=0}\frac{\left[  -k_{35}\right]  _{m_{34}}}{m_{34}!}\nonumber\\
&  \times\frac{\Gamma\left(  k_{12}+1+m_{23}+m_{24}\right)  \Gamma\left(
k_{23}+1\right)  }{\Gamma\left(  k_{12}+k_{23}+2+m_{23}+m_{24}\right)
}\nonumber\\
&  \times\frac{\Gamma\left(  k_{123}+2+m_{23}+m_{24}+m_{34}\right)
\Gamma\left(  k_{34}+1\right)  }{\Gamma\left(  k_{123}+k_{34}+3+m_{23}%
+m_{24}+m_{34}\right)  }\nonumber\\
&  \times\frac{\left(  k_{1234}\right)  ^{m_{24}+m_{34}}}{\left(
k_{1234}+k_{23}+4\right)  _{m_{24}+m_{34}}}\frac{\Gamma\left(  k_{1234}%
+3\right)  \Gamma\left(  k_{45}+1\right)  }{\Gamma\left(  k_{1234}%
+k_{23}+4\right)  }.
\end{align}
Finally, the ratios of the $7$-point RSSA can be easily calculated to be%
\begin{align}
\frac{A^{\left\{  p_{1},p_{2},p_{3}\right\}  ,0,0}}{A^{\left\{  0,0,0\right\}
,0,0}}  &  =\left(  k_{5}^{T_{1}}\right)  ^{p_{1}}\left(  k_{5}^{T_{2}%
}\right)  ^{p_{2}}\left(  k_{5}^{T_{3}}\right)  ^{p_{3}}\nonumber\\
&  =\left(  \cos\phi_{2}^{5}\right)  ^{p_{1}}\left(  \sin\phi_{2}^{5}\cos
\phi_{3}^{5}\right)  ^{p_{2}}\left(  \sin\phi_{2}^{5}\sin\phi_{3}^{5}\right)
^{p_{3}}\nonumber\\
&  =\left(  \cos\theta_{1}\right)  ^{p_{1}}\left(  \sin\theta_{1}\cos
\theta_{2}\right)  ^{p_{2}}\left(  \sin\theta_{1}\sin\theta_{2}\right)
^{p_{3}}\nonumber\\
&  =\left(  \omega_{1}\right)  ^{p_{1}}\left(  \omega_{2}\right)  ^{p_{2}%
}\left(  \omega_{3}\right)  ^{p_{3}},
\end{align}
which is the same as Eq.(\ref{100}) with $m=q=0$ and $r=3$.

\subsection{The $7$-point Regge stringy scaling}

In this section we calculate the $7$-point RSSA with $r=4$ in Eq.(\ref{rr}).
The kinematics are%
\begin{align}
k_{1}  &  =\left(  \sqrt{p^{2}+M_{1}^{2}},-p,0,0,0,0\right)  ,\nonumber\\
k_{2}  &  =\left(  \sqrt{p^{2}+M_{2}^{2}},p,0,0,0,0\right)  ,\nonumber\\
k_{3}  &  =\left(  -\sqrt{q_{3}^{2}+M_{3}^{2}},-q_{3}\cos\phi_{1}^{3}%
,-q_{3}\sin\phi_{1}^{3},0,0,0\right)  ,\nonumber\\
k_{4}  &  =\left(  -\sqrt{q_{4}^{2}+M_{4}^{2}},-q_{4}\cos\phi_{1}^{4}%
,-q_{4}\sin\phi_{1}^{4}\cos\phi_{2}^{4},-q_{4}\sin\phi_{1}^{4}\sin\phi_{2}%
^{4},0,0\right)  ,\nonumber\\
k_{5}  &  =\left(  -\sqrt{q_{5}^{2}+M_{5}^{2}},-q_{5}\cos\phi_{1}^{5}%
,-q_{5}\sin\phi_{1}^{5}\cos\phi_{2}^{5},-q_{5}\sin\phi_{1}^{5}\sin\phi_{2}%
^{5}\cos\phi_{3}^{5},-q_{5}\sin\phi_{1}^{5}\sin\phi_{2}^{5}\sin\phi_{3}%
^{5},0\right)  ,\nonumber\\
k_{6}  &  =\left(
\begin{array}
[c]{c}%
-\sqrt{q_{5}^{2}+M_{5}^{2}},-q_{6}\cos\phi_{1}^{6},-q_{6}\sin\phi_{1}^{6}%
\cos\phi_{2}^{6},-q_{6}\sin\phi_{1}^{6}\sin\phi_{2}^{6}\cos\phi_{3}^{6},\\
-q_{6}\sin\phi_{1}^{6}\sin\phi_{2}^{6}\sin\phi_{3}^{6}\cos\phi_{4}^{6}%
,-q_{6}\sin\phi_{1}^{6}\sin\phi_{2}^{6}\sin\phi_{3}^{6}\sin\phi_{4}^{6}%
\end{array}
\right)  ,\nonumber\\
k_{7}  &  =\left(
\begin{array}
[c]{c}%
-\sqrt{q_{5}^{2}+M_{5}^{2}},-q_{7}\cos\phi_{1}^{7},-q_{7}\sin\phi_{1}^{7}%
\cos\phi_{2}^{7},-q_{7}\sin\phi_{1}^{7}\sin\phi_{2}^{7}\cos\phi_{3}^{7},\\
-q_{7}\sin\phi_{1}^{7}\sin\phi_{2}^{7}\sin\phi_{3}^{7}\cos\phi_{4}^{7}%
,-q_{7}\sin\phi_{1}^{7}\sin\phi_{2}^{7}\sin\phi_{3}^{7}\sin\phi_{4}^{7}%
\end{array}
\right)  .
\end{align}
The tensor state we are going to consider is%
\begin{equation}
\left(  \alpha_{-1}^{T_{1}}\right)  ^{N+p_{1}}\left(  \alpha_{-1}^{T_{2}%
}\right)  ^{p_{2}}\left(  \alpha_{-1}^{T_{3}}\right)  ^{p_{3}}\left(
\alpha_{-1}^{T_{4}}\right)  ^{p_{4}}\left\vert 0,k\right\rangle ,p_{1}%
+p_{2}+p_{3}+p_{4}=0.
\end{equation}
We will use the notation defined in Eq.(\ref{kk}), so we have the following
$\frac{7\left(  7-3\right)  }{2}=14$ independent kinematics variables%
\begin{equation}
k_{12},k_{13},k_{14},k_{15},k_{23},k_{24},k_{25},k_{26},k_{34},k_{35}%
,k_{36},k_{45},k_{46},k_{56}.
\end{equation}
The RSSA of one tensor state and $6$ tachyon states is
\begin{align}
A^{\left\{  p_{1},p_{2},p_{3},p_{4}\right\}  ,0,0}  &  =\int_{0}^{1}dx_{5}%
\int_{0}^{x_{5}}dx_{4}\int_{0}^{x_{4}}dx_{3}\int_{0}^{x_{3}}dx_{2}\cdot
x_{2}^{k_{12}}x_{3}^{k_{13}}x_{4}^{k_{14}}x_{5}^{k_{15}}\left(  x_{3}%
-x_{2}\right)  ^{k_{23}}\left(  x_{4}-x_{2}\right)  ^{k_{24}}\left(
x_{5}-x_{2}\right)  ^{k_{25}}\left(  1-x_{2}\right)  ^{k_{26}}\nonumber\\
&  \times\left(  x_{4}-x_{3}\right)  ^{k_{34}}\left(  x_{5}-x_{3}\right)
^{k_{35}}\left(  1-x_{3}\right)  ^{k_{36}}\left(  x_{5}-x_{4}\right)
^{k_{45}}\left(  1-x_{4}\right)  ^{k_{46}}\left(  1-x_{5}\right)  ^{k_{56}%
}\nonumber\\
&  \times\left[  \frac{k_{3}^{T_{1}}}{x_{3}-x_{2}}+\frac{k_{4}^{T_{1}}}%
{x_{4}-x_{2}}+\frac{k_{5}^{T_{1}}}{x_{5}-x_{2}}+\frac{k_{6}^{T_{1}}}{1-x_{2}%
}\right]  ^{N+p_{1}}\nonumber\\
&  \times\left[  \frac{k_{4}^{T_{2}}}{x_{4}-x_{2}}+\frac{k_{5}^{T_{2}}}%
{x_{5}-x_{2}}+\frac{k_{6}^{T_{2}}}{1-x_{2}}\right]  ^{p_{2}}\nonumber\\
&  \times\left[  \frac{k_{5}^{T_{3}}}{x_{5}-x_{2}}+\frac{k_{6}^{T_{3}}%
}{1-x_{2}}\right]  ^{p_{3}}\left[  \frac{k_{6}^{T_{4}}}{1-x_{2}}\right]
^{p_{4}}.
\end{align}
Note that $k_{3}^{T_{2}},k_{3}^{T_{3}},k_{3}^{T_{4}},k_{4}^{T_{3}}%
,k_{4}^{T_{4}},k_{5}^{T_{4}}$ are all zeros. Let us make the following change
of variables%
\begin{equation}
x_{2}=z_{2}z_{3}z_{4}z_{5},x_{3}=z_{3}z_{4}z_{5},x_{4}=z_{4}z_{5},x_{5}%
=z_{5},x_{6}=z_{6}=1
\end{equation}
to obtain%
\begin{align}
A^{\left\{  p_{1},p_{2},p_{3},p_{4}\right\}  ,0,0}  &  =\int_{0}^{1}dz_{5}%
\int_{0}^{1}dz_{4}\int_{0}^{1}dz_{3}\int_{0}^{1}dz_{2}\cdot\left(  z_{3}%
z_{4}^{2}z_{5}^{3}\right)  \left(  z_{2}z_{3}z_{4}z_{5}\right)  ^{k_{12}%
}\left(  z_{3}z_{4}z_{5}\right)  ^{k_{13}}\left(  z_{4}z_{5}\right)  ^{k_{14}%
}z_{5}^{k_{15}}\nonumber\\
&  \times\left(  z_{3}z_{4}z_{5}-z_{2}z_{3}z_{4}z_{5}\right)  ^{k_{23}}\left(
z_{4}z_{5}-z_{2}z_{3}z_{4}z_{5}\right)  ^{k_{24}}\left(  z_{5}-z_{2}z_{3}%
z_{4}z_{5}\right)  ^{k_{25}}\left(  1-z_{2}z_{3}z_{4}z_{5}\right)  ^{k_{26}%
}\nonumber\\
&  \times\left(  z_{4}z_{5}-z_{3}z_{4}z_{5}\right)  ^{k_{34}}\left(
z_{5}-z_{3}z_{4}z_{5}\right)  ^{k_{35}}\left(  1-z_{3}z_{4}z_{5}\right)
^{k_{36}}\left(  z_{5}-z_{4}z_{5}\right)  ^{k_{45}}\left(  1-z_{4}%
z_{5}\right)  ^{k_{46}}\left(  1-z_{5}\right)  ^{k_{56}}\nonumber\\
&  \times\left[  \frac{k_{3}^{T_{1}}}{z_{3}z_{4}z_{5}-z_{2}z_{3}z_{4}z_{5}%
}+\frac{k_{4}^{T_{1}}}{z_{4}z_{5}-z_{2}z_{3}z_{4}z_{5}}+\frac{k_{5}^{T_{1}}%
}{z_{5}-z_{2}z_{3}z_{4}z_{5}}+\frac{k_{6}^{T_{1}}}{1-z_{2}z_{3}z_{4}z_{5}%
}\right]  ^{N+p_{1}}\nonumber\\
&  \times\left[  \frac{k_{4}^{T_{2}}}{z_{4}z_{5}-z_{2}z_{3}z_{4}z_{5}}%
+\frac{k_{5}^{T_{2}}}{z_{5}-z_{2}z_{3}z_{4}z_{5}}+\frac{k_{6}^{T_{2}}}%
{1-z_{2}z_{3}z_{4}z_{5}}\right]  ^{p_{2}}\nonumber\\
&  \times\left[  \frac{k_{5}^{T_{3}}}{z_{5}-z_{2}z_{3}z_{4}z_{5}}+\frac
{k_{6}^{T_{3}}}{1-z_{2}z_{3}z_{4}z_{5}}\right]  ^{p_{3}}\left[  \frac
{k_{6}^{T_{4}}}{1-z_{2}z_{3}z_{4}z_{5}}\right]  ^{p_{4}}.
\end{align}
We use the definition in Eq.(\ref{kk2}) to obtain%
\begin{equation}
k_{123}=k_{12}+k_{13}+k_{23},k_{1234}=k_{123}+k_{14}+k_{24}+k_{34}%
,k_{12345}=k_{1234}+k_{15}+k_{25}+k_{35}+k_{45}.
\end{equation}

After some calculation, we get%
\begin{align}
A^{\left\{  p_{1},p_{2},p_{3},p_{4}\right\}  ,0,0}  &  =\int_{0}^{1}dz_{5}%
\int_{0}^{1}dz_{4}\int_{0}^{1}dz_{3}\int_{0}^{1}dz_{2}\cdot z_{2}^{k_{12}%
}z_{3}^{k_{123}+1}z_{4}^{k_{1234}+2}z_{5}^{k_{12345}+3}\nonumber\\
&  \times\left(  1-z_{2}\right)  ^{k_{23}}\left(  1-z_{2}z_{3}\right)
^{k_{24}}\left(  1-z_{2}z_{3}z_{4}\right)  ^{k_{25}}\left(  1-z_{2}z_{3}%
z_{4}z_{5}\right)  ^{k_{26}}\nonumber\\
&  \times\left(  1-z_{3}\right)  ^{k_{34}}\left(  1-z_{3}z_{4}\right)
^{k_{35}}\left(  1-z_{3}z_{4}z_{5}\right)  ^{k_{36}}\nonumber\\
&  \times\left(  1-z_{4}\right)  ^{k_{45}}\left(  1-z_{4}z_{5}\right)
^{k_{46}}\nonumber\\
&  \times\left(  1-z_{5}\right)  ^{k_{56}}\nonumber\\
&  \times\left[  \frac{k_{3}^{T_{1}}}{z_{3}z_{4}z_{5}-z_{2}z_{3}z_{4}z_{5}%
}+\frac{k_{4}^{T_{1}}}{z_{4}z_{5}-z_{2}z_{3}z_{4}z_{5}}+\frac{k_{5}^{T_{1}}%
}{z_{5}-z_{2}z_{3}z_{4}z_{5}}+\frac{k_{6}^{T_{1}}}{1-z_{2}z_{3}z_{4}z_{5}%
}\right]  ^{N+p_{1}}\nonumber\\
&  \times\left[  \frac{k_{4}^{T_{2}}}{z_{4}z_{5}-z_{2}z_{3}z_{4}z_{5}}%
+\frac{k_{5}^{T_{2}}}{z_{5}-z_{2}z_{3}z_{4}z_{5}}+\frac{k_{6}^{T_{2}}}%
{1-z_{2}z_{3}z_{4}z_{5}}\right]  ^{p_{2}}\nonumber\\
&  \times\left[  \frac{k_{5}^{T_{3}}}{z_{5}-z_{2}z_{3}z_{4}z_{5}}+\frac
{k_{6}^{T_{3}}}{1-z_{2}z_{3}z_{4}z_{5}}\right]  ^{p_{3}}\left[  \frac
{k_{6}^{T_{4}}}{1-z_{2}z_{3}z_{4}z_{5}}\right]  ^{p_{4}}.
\end{align}
The next step is to expand the brackets to get%
\begin{align}
&  A^{\left\{  p_{1},p_{2},p_{3},p_{4}\right\}  ,0,0}\nonumber\\
&  =\int_{0}^{1}dz_{5}\int_{0}^{1}dz_{4}\int_{0}^{1}dz_{3}\int_{0}^{1}%
dz_{2}\cdot z_{2}^{k_{12}}z_{3}^{k_{123}+1}z_{4}^{k_{1234}+2}z_{5}%
^{k_{12345}+3}\nonumber\\
&  \times\left(  1-z_{2}\right)  ^{k_{23}}\left(  1-z_{2}z_{3}\right)
^{k_{24}}\left(  1-z_{2}z_{3}z_{4}\right)  ^{k_{25}}\left(  1-z_{2}z_{3}%
z_{4}z_{5}\right)  ^{k_{26}}\nonumber\\
&  \times\left(  1-z_{3}\right)  ^{k_{34}}\left(  1-z_{3}z_{4}\right)
^{k_{35}}\left(  1-z_{3}z_{4}z_{5}\right)  ^{k_{36}}\nonumber\\
&  \times\left(  1-z_{4}\right)  ^{k_{45}}\left(  1-z_{4}z_{5}\right)
^{k_{46}}\nonumber\\
&  \times\left(  1-z_{5}\right)  ^{k_{56}}\nonumber\\
&  \times\sum_{J_{1}^{1}+J_{2}^{1}+J_{3}^{1}+J_{4}^{1}=N+p_{1}}^{N+p_{1}}%
\frac{\left(  N+p_{1}\right)  !}{J_{1}^{1}!J_{2}^{1}!J_{3}^{1}!J_{4}^{1}%
!}\left(  \frac{k_{3}^{T_{1}}}{z_{3}z_{4}z_{5}-z_{2}z_{3}z_{4}z_{5}}\right)
^{J_{1}^{1}}\left(  \frac{k_{4}^{T_{1}}}{z_{4}z_{5}-z_{2}z_{3}z_{4}z_{5}%
}\right)  ^{J_{2}^{1}}\left(  \frac{k_{5}^{T_{1}}}{z_{5}-z_{2}z_{3}z_{4}z_{5}%
}\right)  ^{J_{3}^{1}}\left(  \frac{k_{6}^{T_{1}}}{1-z_{2}z_{3}z_{4}z_{5}%
}\right)  ^{J_{4}^{1}}\nonumber\\
&  \times\sum_{J_{1}^{2}+J_{2}^{2}+J_{3}^{2}=p_{2}}^{p_{2}}\frac{P_{2}!}%
{J_{1}^{2}!J_{2}^{2}!J_{3}^{2}!}\left(  \frac{k_{4}^{T_{2}}}{z_{4}z_{5}%
-z_{2}z_{3}z_{4}z_{5}}\right)  ^{J_{1}^{2}}\left(  \frac{k_{5}^{T_{2}}}%
{z_{5}-z_{2}z_{3}z_{4}z_{5}}\right)  ^{J_{2}^{2}}\left(  \frac{k_{6}^{T_{2}}%
}{1-z_{2}z_{3}z_{4}z_{5}}\right)  ^{J_{3}^{2}}\nonumber\\
&  \times\sum_{J_{1}^{3}+J_{2}^{3}=p_{3}}^{p_{2}}\frac{P_{3}!}{J_{1}^{3}%
!J_{2}^{3}!}\left(  \frac{k_{5}^{T_{3}}}{z_{5}-z_{2}z_{3}z_{4}z_{5}}\right)
^{J_{1}^{3}}\left(  \frac{k_{6}^{T_{3}}}{1-z_{2}z_{3}z_{4}z_{5}}\right)
^{J_{2}^{3}}\nonumber\\
&  \times\left(  \frac{k_{6}^{T_{4}}}{1-z_{2}z_{3}z_{4}z_{5}}\right)  ^{p_{4}}%
\end{align}
where $J_{1}^{1},J_{2}^{1},J_{3}^{1},J_{4}^{1},J_{1}^{2},J_{2}^{2},J_{3}%
^{2},J_{1}^{3},J_{2}^{3}$ are non-negative integers with $J_{1}^{1}+J_{2}%
^{1}+J_{3}^{1}+J_{4}^{1}=N+p_{1},$ $J_{1}^{2}+J_{2}^{2}+J_{3}^{2}=p_{2}$ and
$J_{1}^{3}+J_{2}^{3}=p_{3}$. Let us rearrange the above equation as%
\begin{align}
A^{\left\{  p_{1},p_{2},p_{3},p_{4}\right\}  ,0,0}  &  =\sum_{J_{1}^{1}%
+J_{2}^{1}+J_{3}^{1}+J_{4}^{1}=N+p_{1}}^{N+p_{1}}\frac{\left(  N+p_{1}\right)
!}{J_{1}^{1}!J_{2}^{1}!J_{3}^{1}!J_{4}^{1}!}\left(  k_{3}^{T_{1}}\right)
^{J_{1}^{1}}\left(  k_{4}^{T_{1}}\right)  ^{J_{2}^{1}}\left(  k_{5}^{T_{1}%
}\right)  ^{J_{3}^{1}}\left(  k_{6}^{T_{1}}\right)  ^{J_{4}^{1}}\nonumber\\
&  \times\sum_{J_{1}^{2}+J_{2}^{2}+J_{3}^{2}=p_{2}}^{p_{2}}\frac{P_{2}!}%
{J_{1}^{2}!J_{2}^{2}!J_{3}^{2}!}\left(  k_{4}^{T_{2}}\right)  ^{J_{1}^{2}%
}\left(  k_{5}^{T_{2}}\right)  ^{J_{2}^{2}}\left(  k_{6}^{T_{2}}\right)
^{J_{3}^{2}}\nonumber\\
&  \times\sum_{J_{1}^{3}+J_{2}^{3}=p_{3}}^{p_{2}}\frac{P_{3}!}{J_{1}^{3}%
!J_{2}^{3}!}\left(  k_{5}^{T_{3}}\right)  ^{J_{1}^{3}}\left(  k_{6}^{T_{3}%
}\right)  ^{J_{2}^{3}}\left(  k_{6}^{T_{4}}\right)  ^{p_{4}}.\nonumber\\
&  \times\int_{0}^{1}dz_{5}\int_{0}^{1}dz_{4}\int_{0}^{1}dz_{3}\int_{0}%
^{1}dz_{2}\cdot z_{2}^{k_{12}}z_{3}^{k_{123}+1}z_{4}^{k_{1234}+2}%
z_{5}^{k_{12345}+3}\nonumber\\
&  \times\left(  1-z_{2}\right)  ^{k_{23}}\left(  1-z_{2}z_{3}\right)
^{k_{24}}\left(  1-z_{2}z_{3}z_{4}\right)  ^{k_{25}}\left(  1-z_{2}z_{3}%
z_{4}z_{5}\right)  ^{k_{26}}\nonumber\\
&  \times\left(  1-z_{3}\right)  ^{k_{34}}\left(  1-z_{3}z_{4}\right)
^{k_{35}}\left(  1-z_{3}z_{4}z_{5}\right)  ^{k_{36}}\nonumber\\
&  \times\left(  1-z_{4}\right)  ^{k_{45}}\left(  1-z_{4}z_{5}\right)
^{k_{46}}\nonumber\\
&  \times\left(  1-z_{5}\right)  ^{k_{56}}\nonumber\\
&  \times\left(  z_{3}z_{4}z_{5}-z_{2}z_{3}z_{4}z_{5}\right)  ^{-\left(
J_{1}^{1}\right)  }\left(  z_{4}z_{5}-z_{2}z_{3}z_{4}z_{5}\right)  ^{-\left(
J_{2}^{1}+J_{1}^{2}\right)  }\nonumber\\
&  \times\left(  z_{5}-z_{2}z_{3}z_{4}z_{5}\right)  ^{-\left(  J_{3}^{1}%
+J_{2}^{2}+J_{1}^{3}\right)  }\left(  1-z_{2}z_{3}z_{4}z_{5}\right)
^{-\left(  J_{4}^{1}+J_{3}^{2}+J_{2}^{3}+p_{4}\right)  },
\end{align}
which means%
\begin{align}
A^{\left\{  p_{1},p_{2},p_{3},p_{4}\right\}  ,0,0}  &  =\sum_{J_{1}^{1}%
+J_{2}^{1}+J_{3}^{1}+J_{4}^{1}=N+p_{1}}^{N+p_{1}}\frac{\left(  N+p_{1}\right)
!}{J_{1}^{1}!J_{2}^{1}!J_{3}^{1}!J_{4}^{1}!}\left(  k_{3}^{T_{1}}\right)
^{J_{1}^{1}}\left(  k_{4}^{T_{1}}\right)  ^{J_{2}^{1}}\left(  k_{5}^{T_{1}%
}\right)  ^{J_{3}^{1}}\left(  k_{6}^{T_{1}}\right)  ^{J_{4}^{1}}\nonumber\\
&  \times\sum_{J_{1}^{2}+J_{2}^{2}+J_{3}^{2}=p_{2}}^{p_{2}}\frac{P_{2}!}%
{J_{1}^{2}!J_{2}^{2}!J_{3}^{2}!}\left(  k_{4}^{T_{2}}\right)  ^{J_{1}^{2}%
}\left(  k_{5}^{T_{2}}\right)  ^{J_{2}^{2}}\left(  k_{6}^{T_{2}}\right)
^{J_{3}^{2}}\nonumber\\
&  \times\sum_{J_{1}^{3}+J_{2}^{3}=p_{3}}^{p_{2}}\frac{P_{3}!}{J_{1}^{3}%
!J_{2}^{3}!}\left(  k_{5}^{T_{3}}\right)  ^{J_{1}^{3}}\left(  k_{6}^{T_{3}%
}\right)  ^{J_{2}^{3}}\left(  k_{6}^{T_{4}}\right)  ^{p_{4}}.\nonumber\\
&  \times\int_{0}^{1}dz_{5}\int_{0}^{1}dz_{4}\int_{0}^{1}dz_{3}\int_{0}%
^{1}dz_{2}\cdot z_{2}^{k_{12}}z_{3}^{k_{123}+1-\left(  J_{1}^{1}\right)
}z_{4}^{k_{1234}+2-\left(  J_{1}^{1}+J_{2}^{1}+J_{1}^{2}\right)  }%
z_{5}^{k_{12345}+3-\left(  J_{1}^{1}+J_{2}^{1}+J_{1}^{2}+J_{3}^{1}+J_{2}%
^{2}+J_{1}^{3}\right)  }\nonumber\\
&  \times\left(  1-z_{2}\right)  ^{k_{23}-J_{1}^{1}}\left(  1-z_{2}%
z_{3}\right)  ^{k_{24}-\left(  J_{2}^{1}+J_{1}^{2}\right)  }\left(
1-z_{2}z_{3}z_{4}\right)  ^{k_{25}-\left(  J_{3}^{1}+J_{2}^{2}+J_{1}%
^{3}\right)  }\left(  1-z_{2}z_{3}z_{4}z_{5}\right)  ^{k_{26}-\left(
J_{4}^{1}+J_{3}^{2}+J_{2}^{3}+p_{4}\right)  }\nonumber\\
&  \times\left(  1-z_{3}\right)  ^{k_{34}}\left(  1-z_{3}z_{4}\right)
^{k_{35}}\left(  1-z_{3}z_{4}z_{5}\right)  ^{k_{36}}\nonumber\\
&  \times\left(  1-z_{4}\right)  ^{k_{45}}\left(  1-z_{4}z_{5}\right)
^{k_{46}}\nonumber\\
&  \times\left(  1-z_{5}\right)  ^{k_{56}}.
\end{align}
Then we expand the crossing terms to get%
\begin{align}
&  A^{\left\{  p_{1},p_{2},p_{3},p_{4}\right\}  ,0,0}\nonumber\\
&  =\sum_{J_{1}^{1}+J_{2}^{1}+J_{3}^{1}+J_{4}^{1}=N+p_{1}}^{N+p_{1}}%
\frac{\left(  N+p_{1}\right)  !}{J_{1}^{1}!J_{2}^{1}!J_{3}^{1}!J_{4}^{1}%
!}\left(  k_{3}^{T_{1}}\right)  ^{J_{1}^{1}}\left(  k_{4}^{T_{1}}\right)
^{J_{2}^{1}}\left(  k_{5}^{T_{1}}\right)  ^{J_{3}^{1}}\left(  k_{6}^{T_{1}%
}\right)  ^{J_{4}^{1}}\nonumber\\
&  \times\sum_{J_{1}^{2}+J_{2}^{2}+J_{3}^{2}=p_{2}}^{p_{2}}\frac{P_{2}!}%
{J_{1}^{2}!J_{2}^{2}!J_{3}^{2}!}\left(  k_{4}^{T_{2}}\right)  ^{J_{1}^{2}%
}\left(  k_{5}^{T_{2}}\right)  ^{J_{2}^{2}}\left(  k_{6}^{T_{2}}\right)
^{J_{3}^{2}}\nonumber\\
&  \times\sum_{J_{1}^{3}+J_{2}^{3}=p_{3}}^{p_{2}}\frac{P_{3}!}{J_{1}^{3}%
!J_{2}^{3}!}\left(  k_{5}^{T_{3}}\right)  ^{J_{1}^{3}}\left(  k_{6}^{T_{3}%
}\right)  ^{J_{2}^{3}}\left(  k_{6}^{T_{4}}\right)  ^{p_{4}}.\nonumber\\
&  \times\int_{0}^{1}dz_{5}\int_{0}^{1}dz_{4}\int_{0}^{1}dz_{3}\int_{0}%
^{1}dz_{2}\cdot z_{2}^{k_{12}}z_{3}^{k_{123}+1-\left(  J_{1}^{1}\right)
}z_{4}^{k_{1234}+2-\left(  J_{1}^{1}+J_{2}^{1}+J_{1}^{2}\right)  }%
z_{5}^{k_{12345}+3-\left(  J_{1}^{1}+J_{2}^{1}+J_{1}^{2}+J_{3}^{1}+J_{2}%
^{2}+J_{1}^{3}\right)  }\nonumber\\
&  \times\left(  1-z_{2}\right)  ^{k_{23}-J_{1}^{1}}\left(  1-z_{3}\right)
^{k_{34}}\left(  1-z_{4}\right)  ^{k_{45}}\left(  1-z_{5}\right)  ^{k_{56}%
}\nonumber\\
&  \times\sum_{m_{23}}\frac{\left[  -k_{24}+\left(  J_{2}^{1}+J_{1}%
^{2}\right)  \right]  _{m_{23}}}{m_{23}!}\left(  z_{2}z_{3}\right)  ^{m_{23}%
}\sum_{m_{24}}\frac{\left[  -k_{25}+\left(  J_{3}^{1}+J_{2}^{2}+J_{1}%
^{3}\right)  \right]  _{m_{24}}}{m_{24}!}\left(  z_{2}z_{3}z_{4}\right)
^{m_{24}}\nonumber\\
&  \times\sum_{m_{25}}\frac{\left[  -k_{26}+\left(  J_{4}^{1}+J_{3}^{2}%
+J_{2}^{3}+p_{4}\right)  \right]  _{m_{25}}}{m_{25}!}\left(  z_{2}z_{3}%
z_{4}z_{5}\right)  ^{m_{25}}\nonumber\\
&  \times\sum_{m_{34}}\frac{\left[  -k_{35}\right]  _{m_{34}}}{m_{34}!}\left(
z_{3}z_{4}\right)  ^{m_{34}}\sum_{m_{35}}\frac{\left[  -k_{36}\right]
_{m_{34}}}{m_{35}!}\left(  z_{3}z_{4}z_{5}\right)  ^{m_{35}}\sum_{m_{45}}%
\frac{\left[  -k_{46}\right]  _{m_{34}}}{m_{45}!}\left(  z_{4}z_{5}\right)
^{m_{45}},
\end{align}
which gives%
\begin{align}
&  A^{\left\{  p_{1},p_{2},p_{3},p_{4}\right\}  ,0,0}\nonumber\\
&  =\sum_{J_{1}^{1}+J_{2}^{1}+J_{3}^{1}+J_{4}^{1}=N+p_{1}}^{N+p_{1}}%
\frac{\left(  N+p_{1}\right)  !}{J_{1}^{1}!J_{2}^{1}!J_{3}^{1}!J_{4}^{1}%
!}\left(  k_{3}^{T_{1}}\right)  ^{J_{1}^{1}}\left(  k_{4}^{T_{1}}\right)
^{J_{2}^{1}}\left(  k_{5}^{T_{1}}\right)  ^{J_{3}^{1}}\left(  k_{6}^{T_{1}%
}\right)  ^{J_{4}^{1}}\nonumber\\
&  \times\sum_{J_{1}^{2}+J_{2}^{2}+J_{3}^{2}=p_{2}}^{p_{2}}\frac{P_{2}!}%
{J_{1}^{2}!J_{2}^{2}!J_{3}^{2}!}\left(  k_{4}^{T_{2}}\right)  ^{J_{1}^{2}%
}\left(  k_{5}^{T_{2}}\right)  ^{J_{2}^{2}}\left(  k_{6}^{T_{2}}\right)
^{J_{3}^{2}}\nonumber\\
&  \times\sum_{J_{1}^{3}+J_{2}^{3}=p_{3}}^{p_{2}}\frac{P_{3}!}{J_{1}^{3}%
!J_{2}^{3}!}\left(  k_{5}^{T_{3}}\right)  ^{J_{1}^{3}}\left(  k_{6}^{T_{3}%
}\right)  ^{J_{2}^{3}}\left(  k_{6}^{T_{4}}\right)  ^{p_{4}}\nonumber\\
&  \times\sum_{m_{23}}\frac{\left[  -k_{24}+\left(  J_{2}^{1}+J_{1}%
^{2}\right)  \right]  _{m_{23}}}{m_{23}!}\sum_{m_{24}}\frac{\left[
-k_{25}+\left(  J_{3}^{1}+J_{2}^{2}+J_{1}^{3}\right)  \right]  _{m_{24}}%
}{m_{24}!}\sum_{m_{25}}\frac{\left[  -k_{26}+\left(  J_{4}^{1}+J_{3}^{2}%
+J_{2}^{3}+p_{4}\right)  \right]  _{m_{25}}}{m_{25}!}\nonumber\\
&  \times\sum_{m_{34}}\frac{\left[  -k_{35}\right]  _{m_{34}}}{m_{34}!}%
\sum_{m_{35}}\frac{\left[  -k_{36}\right]  _{m_{35}}}{m_{35}!}\sum_{m_{45}%
}\frac{\left[  -k_{46}\right]  _{m_{45}}}{m_{45}!}\nonumber\\
&  \times\int_{0}^{1}dz_{2}\cdot z_{2}^{k_{12}+m_{23}+m_{24}+m_{25}}\left(
1-z_{2}\right)  ^{k_{23}-J_{1}^{1}}\nonumber\\
&  \times\int_{0}^{1}dz_{3}\cdot z_{3}^{k_{123}+1-\left(  J_{1}^{1}\right)
+m_{23}+m_{24}+m_{25}+m_{34}+m_{35}}\left(  1-z_{3}\right)  ^{k_{34}%
}\nonumber\\
&  \times\int_{0}^{1}dz_{4}\cdot z_{4}^{k_{1234}+2-\left(  J_{1}^{1}+J_{2}%
^{1}+J_{1}^{2}\right)  +m_{24}+m_{25}+m_{34}+m_{35}+m_{45}}\left(
1-z_{4}\right)  ^{k_{45}}\nonumber\\
&  \times\int_{0}^{1}dz_{5}z_{5}^{k_{12345}+3-\left(  J_{1}^{1}+J_{2}%
^{1}+J_{1}^{2}+J_{3}^{1}+J_{2}^{2}+J_{1}^{3}\right)  +m_{25}+m_{35}+m_{45}%
}\left(  1-z_{5}\right)  ^{k_{56}}.
\end{align}
After integration, we obtain%
\begin{align}
&  A^{\left\{  p_{1},p_{2},p_{3},p_{4}\right\}  ,0,0}\nonumber\\
&  =\sum_{J_{1}^{1}+J_{2}^{1}+J_{3}^{1}+J_{4}^{1}=N+p_{1}}^{N+p_{1}}%
\frac{\left(  N+p_{1}\right)  !}{J_{1}^{1}!J_{2}^{1}!J_{3}^{1}!J_{4}^{1}%
!}\left(  k_{3}^{T_{1}}\right)  ^{J_{1}^{1}}\left(  k_{4}^{T_{1}}\right)
^{J_{2}^{1}}\left(  k_{5}^{T_{1}}\right)  ^{J_{3}^{1}}\left(  k_{6}^{T_{1}%
}\right)  ^{J_{4}^{1}}\nonumber\\
&  \times\sum_{J_{1}^{2}+J_{2}^{2}+J_{3}^{2}=p_{2}}^{p_{2}}\frac{P_{2}!}%
{J_{1}^{2}!J_{2}^{2}!J_{3}^{2}!}\left(  k_{4}^{T_{2}}\right)  ^{J_{1}^{2}%
}\left(  k_{5}^{T_{2}}\right)  ^{J_{2}^{2}}\left(  k_{6}^{T_{2}}\right)
^{J_{3}^{2}}\nonumber\\
&  \times\sum_{J_{1}^{3}+J_{2}^{3}=p_{3}}^{p_{2}}\frac{P_{3}!}{J_{1}^{3}%
!J_{2}^{3}!}\left(  k_{5}^{T_{3}}\right)  ^{J_{1}^{3}}\left(  k_{6}^{T_{3}%
}\right)  ^{J_{2}^{3}}\left(  k_{6}^{T_{4}}\right)  ^{p_{4}}\nonumber\\
&  \times\sum_{m_{23}}\frac{\left[  -k_{24}+\left(  J_{2}^{1}+J_{1}%
^{2}\right)  \right]  _{m_{23}}}{m_{23}!}\sum_{m_{24}}\frac{\left[
-k_{25}+\left(  J_{3}^{1}+J_{2}^{2}+J_{1}^{3}\right)  \right]  _{m_{24}}%
}{m_{24}!}\sum_{m_{25}}\frac{\left[  -k_{26}+\left(  J_{4}^{1}+J_{3}^{2}%
+J_{2}^{3}+p_{4}\right)  \right]  _{m_{25}}}{m_{25}!}\nonumber\\
&  \times\sum_{m_{34}}\frac{\left[  -k_{35}\right]  _{m_{34}}}{m_{34}!}%
\sum_{m_{35}}\frac{\left[  -k_{36}\right]  _{m_{35}}}{m_{35}!}\sum_{m_{45}%
}\frac{\left[  -k_{46}\right]  _{m_{45}}}{m_{45}!}\nonumber\\
&  \times\frac{\Gamma\left(  k_{12}+1+m_{23}+m_{24}+m_{25}\right)
\Gamma\left(  k_{23}+1-J_{1}^{1}\right)  }{\Gamma\left(  k_{12}+k_{23}%
+2+m_{23}+m_{24}+m_{25}\right)  }\nonumber\\
&  \times\frac{\Gamma\left(  k_{123}+2-J_{1}^{1}+m_{23}+m_{24}+m_{25}%
+m_{34}+m_{35}\right)  \Gamma\left(  k_{34}+1\right)  }{\Gamma\left(
k_{123}+k_{34}+3-J_{1}^{1}+m_{23}+m_{24}+m_{25}+m_{34}+m_{35}\right)
}\nonumber\\
&  \times\frac{\Gamma\left(  k_{1234}+3-\left(  J_{1}^{1}+J_{2}^{1}+J_{1}%
^{2}\right)  +m_{24}+m_{25}+m_{34}+m_{35}+m_{45}\right)  \Gamma\left(
k_{45}+1\right)  }{\Gamma\left(  k_{1234}+k_{45}+4-\left(  J_{1}^{1}+J_{2}%
^{1}+J_{1}^{2}\right)  +m_{24}+m_{25}+m_{34}+m_{35}+m_{45}\right)
}\nonumber\\
&  \times\frac{\Gamma\left(  k_{12345}+4-\left(  J_{1}^{1}+J_{2}^{1}+J_{1}%
^{2}+J_{3}^{1}+J_{2}^{2}+J_{1}^{3}\right)  +m_{25}+m_{35}+m_{45}\right)
\Gamma\left(  k_{56}+1\right)  }{\Gamma\left(  k_{12345}+k_{56}+5-\left(
J_{1}^{1}+J_{2}^{1}+J_{1}^{2}+J_{3}^{1}+J_{2}^{2}+J_{1}^{3}\right)
+m_{25}+m_{35}+m_{45}\right)  }.
\end{align}
Now we choose to work on the Regge regime defined by%
\begin{equation}
k_{12345}\sim s,k_{12345}+k_{56}\sim t.
\end{equation}
In this regime, the RSSA can be approximated as%
\begin{align}
&  A^{\left\{  p_{1},p_{2},p_{3},p_{4}\right\}  ,0,0}\nonumber\\
&  \sim\sum_{J_{1}^{1}+J_{2}^{1}+J_{3}^{1}+J_{4}^{1}=N+p_{1}}^{N+p_{1}}%
\frac{\left(  N+p_{1}\right)  !}{J_{1}^{1}!J_{2}^{1}!J_{3}^{1}!J_{4}^{1}%
!}\left(  k_{3}^{T_{1}}\right)  ^{J_{1}^{1}}\left(  k_{4}^{T_{1}}\right)
^{J_{2}^{1}}\left(  k_{5}^{T_{1}}\right)  ^{J_{3}^{1}}\left(  k_{6}^{T_{1}%
}\right)  ^{J_{4}^{1}}\nonumber\\
&  \times\sum_{J_{1}^{2}+J_{2}^{2}+J_{3}^{2}=p_{2}}^{p_{2}}\frac{P_{2}!}%
{J_{1}^{2}!J_{2}^{2}!J_{3}^{2}!}\left(  k_{4}^{T_{2}}\right)  ^{J_{1}^{2}%
}\left(  k_{5}^{T_{2}}\right)  ^{J_{2}^{2}}\left(  k_{6}^{T_{2}}\right)
^{J_{3}^{2}}\nonumber\\
&  \times\sum_{J_{1}^{3}+J_{2}^{3}=p_{3}}^{p_{2}}\frac{P_{3}!}{J_{1}^{3}%
!J_{2}^{3}!}\left(  k_{5}^{T_{3}}\right)  ^{J_{1}^{3}}\left(  k_{6}^{T_{3}%
}\right)  ^{J_{2}^{3}}\left(  k_{6}^{T_{4}}\right)  ^{p_{4}}\nonumber\\
&  \times\sum_{m_{23}}\frac{\left[  -k_{24}+\left(  J_{2}^{1}+J_{1}%
^{2}\right)  \right]  _{m_{23}}}{m_{23}!}\sum_{m_{24}}\frac{\left[
-k_{25}+\left(  J_{3}^{1}+J_{2}^{2}+J_{1}^{3}\right)  \right]  _{m_{24}}%
}{m_{24}!}\sum_{m_{25}}\frac{\left[  -k_{26}+\left(  J_{4}^{1}+J_{3}^{2}%
+J_{2}^{3}+p_{4}\right)  \right]  _{m_{25}}}{m_{25}!}\nonumber\\
&  \times\sum_{m_{34}}\frac{\left[  -k_{35}\right]  _{m_{34}}}{m_{34}!}%
\sum_{m_{35}}\frac{\left[  -k_{36}\right]  _{m_{35}}}{m_{35}!}\sum_{m_{45}%
}\frac{\left[  -k_{46}\right]  _{m_{45}}}{m_{45}!}\nonumber\\
&  \times\frac{\Gamma\left(  k_{12}+1+m_{23}+m_{24}+m_{25}\right)
\Gamma\left(  k_{23}+1-J_{1}^{1}\right)  }{\Gamma\left(  k_{12}+k_{23}%
+2-J_{1}^{1}+m_{23}+m_{24}+m_{25}\right)  }\nonumber\\
&  \times\frac{\Gamma\left(  k_{123}+2-J_{1}^{1}+m_{23}+m_{24}+m_{25}%
+m_{34}+m_{35}\right)  \Gamma\left(  k_{34}+1\right)  }{\Gamma\left(
k_{123}+k_{34}+3-J_{1}^{1}+m_{23}+m_{24}+m_{25}+m_{34}+m_{35}\right)
}\nonumber\\
&  \times\frac{\Gamma\left(  k_{1234}+3-\left(  J_{1}^{1}+J_{2}^{1}+J_{1}%
^{2}\right)  +m_{24}+m_{25}+m_{34}+m_{35}+m_{45}\right)  \Gamma\left(
k_{45}+1\right)  }{\Gamma\left(  k_{1234}+k_{45}+4-\left(  J_{1}^{1}+J_{2}%
^{1}+J_{1}^{2}\right)  +m_{24}+m_{25}+m_{34}+m_{35}+m_{45}\right)
}\nonumber\\
&  \times\frac{\left(  k_{12345}\right)  ^{-\left(  J_{1}^{1}+J_{2}^{1}%
+J_{1}^{2}+J_{3}^{1}+J_{2}^{2}+J_{1}^{3}\right)  +m_{25}+m_{35}+m_{45}}%
}{\left(  k_{12345}+k_{56}+5\right)  _{-\left(  J_{1}^{1}+J_{2}^{1}+J_{1}%
^{2}+J_{3}^{1}+J_{2}^{2}+J_{1}^{3}\right)  +m_{25}+m_{35}+m_{45}}}\frac
{\Gamma\left(  k_{12345}+4\right)  \Gamma\left(  k_{56}+1\right)  }%
{\Gamma\left(  k_{12345}+k_{56}+5\right)  }.
\end{align}
To get the leading order in $k_{12345}$, we take
\[
J_{1}^{1}=J_{2}^{1}=J_{3}^{1}=J_{1}^{2}=J_{2}^{2}=J_{1}^{3}=0,
\]
which implies%
\begin{equation}
J_{4}^{1}=N+p_{1},J_{3}^{2}=p_{2},J_{2}^{3}=p_{3}.
\end{equation}
With $p_{1}+p_{2}+p_{3}+p_{4}=0$, the leading term is

\bigskip%
\begin{align}
&  A^{\left\{  p_{1},p_{2},p_{3},p_{4}\right\}  ,0,0}\nonumber\\
&  \sim\left(  k_{6}^{T_{1}}\right)  ^{N+p_{1}}\left(  k_{6}^{T_{2}}\right)
^{p_{2}}\left(  k_{6}^{T_{3}}\right)  ^{p_{3}}\left(  k_{6}^{T_{4}}\right)
^{p_{4}}\nonumber\\
&  \times\sum_{m_{23}}\frac{\left[  -k_{24}\right]  _{m_{23}}}{m_{23}!}%
\sum_{m_{24}}\frac{\left[  -k_{25}\right]  _{m_{24}}}{m_{24}!}\sum_{m_{25}%
}\frac{\left[  -k_{26}+N\right]  _{m_{25}}}{m_{25}!}\sum_{m_{34}}\frac{\left[
-k_{35}\right]  _{m_{34}}}{m_{34}!}\sum_{m_{35}}\frac{\left[  -k_{36}\right]
_{m_{35}}}{m_{35}!}\sum_{m_{45}}\frac{\left[  -k_{46}\right]  _{m_{45}}%
}{m_{45}!}\nonumber\\
&  \times\frac{\Gamma\left(  k_{12}+1+m_{23}+m_{24}+m_{25}\right)
\Gamma\left(  k_{23}+1\right)  }{\Gamma\left(  k_{12}+k_{23}+2+m_{23}%
+m_{24}+m_{25}\right)  }\nonumber\\
&  \times\frac{\Gamma\left(  k_{123}+2+m_{23}+m_{24}+m_{25}+m_{34}%
+m_{35}\right)  \Gamma\left(  k_{34}+1\right)  }{\Gamma\left(  k_{123}%
+k_{34}+3+m_{23}+m_{24}+m_{25}+m_{34}+m_{35}\right)  }\nonumber\\
&  \times\frac{\Gamma\left(  k_{1234}+3+m_{24}+m_{25}+m_{34}+m_{35}%
+m_{45}\right)  \Gamma\left(  k_{45}+1\right)  }{\Gamma\left(  k_{1234}%
+k_{45}+4+m_{24}+m_{25}+m_{34}+m_{35}+m_{45}\right)  }\nonumber\\
&  \times\frac{\left(  k_{12345}\right)  ^{+m_{25}+m_{35}+m_{45}}}{\left(
k_{12345}+k_{56}+5\right)  _{m_{25}+m_{35}+m_{45}}}\frac{\Gamma\left(
k_{12345}+4\right)  \Gamma\left(  k_{56}+1\right)  }{\Gamma\left(
k_{12345}+k_{56}+5\right)  }.
\end{align}
So the ratios of the $7$-point RSSA is
\begin{align}
\frac{A^{\left\{  p_{1},p_{2},p_{3},p_{4}\right\}  ,0,0}}{A^{\left\{
0,0,0,0\right\}  ,0,0}}  &  =\left(  k_{6}^{T_{1}}\right)  ^{p_{1}}\left(
k_{6}^{T_{2}}\right)  ^{p_{2}}\left(  k_{6}^{T_{3}}\right)  ^{p_{3}}\left(
k_{6}^{T_{4}}\right)  ^{p_{4}}\nonumber\\
&  =\left(  \cos\phi_{2}^{6}\right)  ^{p_{1}}\left(  \sin\phi_{2}^{6}\cos
\phi_{3}^{6}\right)  ^{p_{2}}\left(  \sin\phi_{2}^{6}\sin\phi_{3}^{6}\cos
\phi_{4}^{6}\right)  ^{p_{3}}\left(  \sin\phi_{2}^{6}\sin\phi_{3}^{6}\sin
\phi_{4}^{6}\right)  ^{p_{4}}\nonumber\\
&  =\left(  \cos\theta_{1}\right)  ^{p_{1}}\left(  \sin\theta_{1}\cos
\theta_{2}\right)  ^{p_{2}}\left(  \sin\theta_{1}\sin\theta_{2}\cos\theta
_{3}\right)  ^{p_{3}}\left(  \sin\theta_{1}\sin\theta_{2}\sin\theta
_{3}\right)  ^{p_{4}}\nonumber\\
&  =\left(  \omega_{1}\right)  ^{p_{1}}\left(  \omega_{2}\right)  ^{p_{2}%
}\left(  \omega_{3}\right)  ^{p_{3}}\left(  \omega_{4}\right)  ^{p_{4}},
\end{align}
\newline which is the same as Eq.(\ref{100}) with $m=q=0$ and $r=4$.%

%TCIMACRO{\TeXButton{equation number}{\setcounter{equation}{0}
%\renewcommand{\theequation}{\arabic{section}.\arabic{equation}}}}%
%BeginExpansion
\setcounter{equation}{0}
\renewcommand{\theequation}{\arabic{section}.\arabic{equation}}%
%EndExpansion

\section{The $n$-point Regge stringy scaling}

In this section, we generalize the previous calculations to the case of
$n$-point RSSA. We first define the $26$-dimensional momenta in the CM frame
to be%
\begin{align}
k_{1} &  =\left(  \sqrt{p^{2}+M_{1}^{2}},-p,0^{r}\right)  ,\nonumber\\
k_{2} &  =\left(  \sqrt{p^{2}+M_{2}^{2}},p,0^{r}\right)  ,\nonumber\\
&  \vdots\nonumber\\
k_{j} &  =\left(  -\sqrt{q_{j}^{2}+M_{j}^{2}},-q_{j}\Omega_{1}^{j}%
,-q_{j}\Omega_{2}^{j},\cdots,-q_{j}\Omega_{r}^{j},-q_{j}\Omega_{r+1}%
^{j}\right)  \label{55}%
\end{align}
where $j=3,4,\cdots,n$, and%
\begin{equation}
\Omega_{i}^{j}=\cos\phi_{i}^{j}\prod\limits_{\sigma=1}^{i-1}\sin\phi_{\sigma
}^{j}\text{ with }\phi_{j-1}^{j}=0,\text{ }\phi_{i>r}^{j}=0\text{ and }%
r\leq\min\left\{  n-3,24\right\}
\end{equation}
are the solid angles in the $\left(  j-2\right)  $-dimensional spherical space
with $\sum_{i=1}^{j-2}\left(  \Omega_{i}^{j}\right)  ^{2}=1$. In
Eq.(\ref{55}), $0^{r}$ denotes the $r$-dimensional null vector. The amplitude
of one tensor state%
\begin{equation}
\left(  \alpha_{-1}^{T_{1}}\right)  ^{N+p_{1}}\left(  \alpha_{-1}^{T_{2}%
}\right)  ^{p_{2}}\cdots\left(  \alpha_{-1}^{T_{r}}\right)  ^{p_{r}}\left\vert
0,k\right\rangle ,p_{1}+p_{2}+\cdots+p_{r}=0
\end{equation}
and $n-1$ tachyon states is%
\begin{align}
&  A^{\left\{  p_{1},p_{2},\cdots,p_{r}\right\}  ,0,0}\nonumber\\
&  =\int_{0}^{1}dx_{n-2}\int_{0}^{x_{n-2}}dx_{n-3}\cdots\int_{0}^{x_{4}}%
dx_{3}\int_{0}^{x_{3}}dx_{2}\times\prod_{0\leq i<j\leq n-1}\left(  x_{j}%
-x_{i}\right)  ^{k_{ij}}\prod_{\sigma=1}^{r}\left[  \sum_{j=\sigma+2}%
^{n-1}\left(  \frac{k_{j}^{T_{\sigma}}}{x_{j}-x_{2}}\right)  \right]
^{\mathcal{P}_{\sigma}}\label{nn}%
\end{align}
where we have defined%
\begin{equation}
\mathcal{P}_{1}=N+p_{1},\mathcal{P}_{\sigma\neq1}=p_{\sigma}.
\end{equation}
Now, let's explicitly write down the second product part of Eq.(\ref{nn}) as
\begin{align}
&  A^{\left\{  p_{1},p_{2},\cdots,p_{r}\right\}  ,0,0}\nonumber\\
&  =\int_{0}^{1}dx_{n-2}\int_{0}^{x_{n-2}}dx_{n-3}\cdots\int_{0}^{x_{4}}%
dx_{3}\int_{0}^{x_{3}}dx_{2}\times\prod_{0\leq i<j\leq n-1}\left(  x_{j}%
-x_{i}\right)  ^{k_{ij}}\nonumber\\
&  \times\left[  \frac{k_{3}^{T_{1}}}{x_{3}-x_{2}}+\frac{k_{4}^{T_{1}}}%
{x_{4}-x_{2}}+\frac{k_{5}^{T_{1}}}{x_{5}-x_{2}}\cdots+\frac{k_{n-1}^{T_{1}}%
}{1-x_{2}}\right]  ^{\mathcal{P}_{1}}\nonumber\\
&  \times\left[  \frac{k_{4}^{T_{2}}}{x_{4}-x_{2}}+\frac{k_{5}^{T_{2}}}%
{x_{5}-x_{2}}+\cdots+\frac{k_{n-1}^{T_{2}}}{1-x_{2}}\right]  ^{\mathcal{P}%
_{2}}\nonumber\\
&  \vdots\nonumber\\
&  \times\left[  \frac{k_{r+2}^{T_{r}}}{x_{r+2}-x_{2}}+\cdots+\frac
{k_{n-1}^{T_{r}}}{1-x_{2}}\right]  ^{\mathcal{P}_{r}}.
\end{align}
For convience, from now on we add trivial terms with $\mathcal{P}_{\sigma}=0$
$(r+1\leq\sigma\leq n-3)$ to the amplitude and obtain%
\begin{align}
&  A^{\left\{  p_{1},p_{2},\cdots,p_{r}\right\}  ,0,0}\nonumber\\
&  =\int_{0}^{1}dx_{n-2}\int_{0}^{x_{n-2}}dx_{n-3}\cdots\int_{0}^{x_{4}}%
dx_{3}\int_{0}^{x_{3}}dx_{2}\times\prod_{0\leq i<j\leq n-1}\left(  x_{j}%
-x_{i}\right)  ^{k_{ij}}\nonumber\\
&  \times\left[  \frac{k_{3}^{T_{1}}}{x_{3}-x_{2}}+\frac{k_{4}^{T_{1}}}%
{x_{4}-x_{2}}+\frac{k_{5}^{T_{1}}}{x_{5}-x_{2}}+\frac{k_{5}^{T_{1}}}%
{x_{6}-x_{2}}+\cdots+\frac{k_{n-1}^{T_{1}}}{1-x_{2}}\right]  ^{\mathcal{P}%
_{1}}\nonumber\\
&  \times\left[  \frac{k_{4}^{T_{2}}}{x_{4}-x_{2}}+\frac{k_{5}^{T_{2}}}%
{x_{5}-x_{2}}+\frac{k_{6}^{T_{2}}}{x_{6}-x_{2}}+\cdots+\frac{k_{n-1}^{T_{2}}%
}{1-x_{2}}\right]  ^{\mathcal{P}_{2}}\nonumber\\
&  \vdots\nonumber\\
&  \times\left[  \frac{k_{r+2}^{T_{r}}}{x_{r+2}-x_{2}}+\frac{k_{r+3}^{T_{r}}%
}{x_{r+3}-x_{2}}+\cdots+\frac{k_{n-1}^{T_{r}}}{1-x_{2}}\right]  ^{\mathcal{P}%
_{r}}\nonumber\\
&  \times\left[  \frac{k_{r+3}^{T_{r+1}}}{x_{r+3}-x_{2}}+\cdots+\frac
{k_{n-1}^{T_{r+1}}}{1-x_{2}}\right]  ^{\mathcal{P}_{r+1}}\nonumber\\
&  \vdots\nonumber\\
&  \times\left[  \frac{k_{n-1}^{T_{n-3}}}{1-x_{2}}\right]  ^{\mathcal{P}%
_{n-3}}.
\end{align}
Now we can expand the brackets%
\begin{align}
&  A^{\left\{  p_{1},p_{2},\cdots,p_{r}\right\}  ,0,0}\nonumber\\
&  =\int_{0}^{1}dx_{n-2}\int_{0}^{x_{n-2}}dx_{n-3}\cdots\int_{0}^{x_{4}}%
dx_{3}\int_{0}^{x_{3}}dx_{2}\times\prod_{0\leq i<j\leq n-1}\left(  x_{j}%
-x_{i}\right)  ^{k_{ij}}\nonumber\\
&  \times\sum_{J_{1}^{1}+J_{2}^{1}+\cdots+J_{n-4}^{1}+J_{n-3}^{1}%
\equiv\mathcal{P}_{1}}\frac{\mathcal{P}_{1}!}{J_{1}^{1}!J_{2}^{1}!\cdots
J_{n-4}^{1}!J_{n-3}^{1}!}\left(  \frac{k_{3}^{T_{1}}}{x_{3}-x_{2}}\right)
^{J_{1}^{1}}\left(  \frac{k_{4}^{T_{1}}}{x_{4}-x_{2}}\right)  ^{J_{2}^{1}%
}\cdots\left(  \frac{k_{n-1}^{T_{1}}}{1-x_{2}}\right)  ^{J_{n-3}^{1}%
}\nonumber\\
&  \times\sum_{J_{1}^{2}+J_{2}^{2}+\cdots+J_{n-5}^{2}+J_{n-4}^{2}%
=\mathcal{P}_{2}}\frac{\mathcal{P}_{2}!}{J_{1}^{2}!J_{2}^{2}!\cdots
J_{n-5}^{2}!J_{n-4}^{2}!}\left(  \frac{k_{4}^{T_{2}}}{x_{4}-x_{2}}\right)
^{J_{1}^{2}}\cdots\left(  \frac{k_{n-1}^{T_{2}}}{1-x_{2}}\right)
^{J_{n-4}^{2}}\nonumber\\
&  \vdots\nonumber\\
\times &  \sum_{J_{1}^{n-3}=\mathcal{P}_{n-3}=0}\frac{\mathcal{P}_{n-3}%
!}{J_{1}^{n-3}!}\left(  \frac{k_{n-1}^{T_{n-3}}}{1-x_{2}}\right)
^{J_{1}^{n-3}}%
\end{align}
where all $J$ are non-negative integers. (Note that all $J_{j}^{\sigma\geq
r+1}=0$ due to $\mathcal{P}_{\sigma\geq r+1}=0.$)

After some rearrangements, we can derive%
\begin{align}
&  A^{\left\{  p_{1},p_{2},\cdots,p_{n-3}\right\}  ,0,0}\nonumber\\
&  =\sum_{J_{1}^{1}+J_{2}^{1}+\cdots+J_{n-4}^{1}+J_{n-3}^{1}\equiv
\mathcal{P}_{1}}\frac{\mathcal{P}_{1}!}{J_{1}^{1}!J_{2}^{1}!\cdots J_{n-4}%
^{1}!J_{n-3}^{1}!}\left(  k_{3}^{T_{1}}\right)  ^{J_{1}^{1}}\left(
k_{4}^{T_{1}}\right)  ^{J_{2}^{1}}\cdots\left(  k_{n-1}^{T_{1}}\right)
^{J_{n-3}^{1}}\nonumber\\
&  \times\sum_{J_{1}^{2}+J_{2}^{2}+\cdots+J_{n-5}^{2}+J_{n-4}^{2}%
=\mathcal{P}_{2}}\frac{\mathcal{P}_{2}!}{J_{1}^{2}!J_{2}^{2}!\cdots
J_{n-5}^{2}!J_{n-4}^{2}!}\left(  k_{4}^{T_{2}}\right)  ^{J_{1}^{2}}\left(
k_{5}^{T_{2}}\right)  ^{J_{2}^{2}}\cdots\left(  k_{n-1}^{T_{2}}\right)
^{J_{n-4}^{2}}\nonumber\\
&  \vdots\nonumber\\
&  \times\sum_{J_{1}^{n-3}=\mathcal{P}_{n-3}=0}\frac{\mathcal{P}_{n-3}!}%
{J_{1}^{n-3}!}\left(  k_{n-1}^{T_{n-3}}\right)  ^{J_{1}^{n-3}}\nonumber\\
&  \times\int_{0}^{1}dx_{n-2}\int_{0}^{x_{n-2}}dx_{n-3}\cdots\int_{0}^{x_{4}%
}dx_{3}\int_{0}^{x_{3}}dx_{2}\times\prod_{0\leq i<j\leq n-1}\left(
x_{j}-x_{i}\right)  ^{k_{ij}}\nonumber\\
&  \times\left(  \frac{1}{x_{3}-x_{2}}\right)  ^{J_{1}^{1}}\left(  \frac
{1}{x_{4}-x_{2}}\right)  ^{J_{2}^{1}}\cdots\left(  \frac{1}{1-x_{2}}\right)
^{J_{n-3}^{1}}\nonumber\\
&  \times\left(  \frac{1}{x_{4}-x_{2}}\right)  ^{J_{1}^{2}}\cdots\left(
\frac{1}{1-x_{2}}\right)  ^{J_{n-4}^{2}}\nonumber\\
&  \vdots\nonumber\\
&  \times\left(  \frac{1}{1-x_{2}}\right)  ^{J_{1}^{n-3}}.
\end{align}

\bigskip We can collect terms with the same sum of subscripts and superscripts
to get%

\begin{align}
&  A^{\left\{  p_{1},p_{2},\cdots,p_{n-3}\right\}  ,0,0}\nonumber\\
&  =\prod_{\sigma=1}^{n-3}\left[  \sum_{\sum_{j=1}^{n-2-\sigma}J_{j}^{\sigma
}=\mathcal{P}_{\sigma}}\left(  \mathcal{P}_{\sigma}!\prod_{j=1}^{n-2-\sigma
}\frac{\left(  k_{j+\sigma+1}^{T_{\sigma}}\right)  ^{J_{j}^{\sigma}}}%
{J_{j}^{\sigma}!}\right)  \right] \\
&  \times\int_{0}^{1}dx_{n-2}\int_{0}^{x_{n-2}}dx_{n-3}\cdots\int_{0}^{x_{4}%
}dx_{3}\int_{0}^{x_{3}}dx_{2}\times\prod_{0\leq i<j\leq n-1}\left(
x_{j}-x_{i}\right)  ^{k_{ij}-\delta_{i2}\left(  J_{j-2}^{1}+\cdots+J_{1}%
^{j-2}\right)  }. \label{Before int}%
\end{align}
To perform the integral of the last line in Eq.(\ref{Before int}), let's do
the following change of variables%
\begin{equation}
x_{i}=\prod_{k=i}^{n-2}z_{k},\text{ or }x_{2}=z_{2}z_{3}\cdots z_{n-2}%
,x_{3}=z_{3}z_{4}\cdots z_{n-2},\cdots,x_{n-2}=z_{n-2},x_{n-1}=z_{n-1}=1
\end{equation}
to make all the integral intervals from $0$ to $1$. The integral becomes%
\begin{align}
&  \int_{0}^{1}dx_{n-2}\int_{0}^{x_{n-2}}dx_{n-3}\cdots\int_{0}^{x_{4}}%
dx_{3}\int_{0}^{x_{3}}dx_{2}\times\prod_{0\leq i<j\leq n-1}\left(  x_{j}%
-x_{i}\right)  ^{k_{ij}-\delta_{i2}\left(  J_{j-2}^{1}+\cdots+J_{1}%
^{j-2}\right)  }\nonumber\\
&  =\int_{0}^{1}dz_{n-2}\cdots\int_{0}^{1}dz_{3}\int_{0}^{1}dz_{2}\prod
_{i=1}^{n-4}\left(  z_{i+2}\right)  ^{i}\prod_{0\leq i<j\leq n-1}\left(
\prod_{k=j}^{n-2}z_{k}-\prod_{k=i}^{n-2}z_{k}\right)  ^{k_{ij}-\delta
_{i2}\left(  J_{j-2}^{1}+\cdots+J_{1}^{j-2}\right)  }\nonumber\\
&  =\int_{0}^{1}dz_{n-2}\cdots\int_{0}^{1}dz_{2}\times\prod_{i=1}^{n-4}\left(
z_{i+2}\right)  ^{i}\prod_{0\leq i<j\leq n-1}\left[  \prod_{k=j}^{n-2}%
z_{k}\left(  1-\prod_{k=i}^{j-1}z_{k}\right)  \right]  ^{k_{ij}-\delta
_{i2}\left(  J_{j-2}^{1}+\cdots+J_{1}^{j-2}\right)  }.
\end{align}
Now, the amplitude can be explicitly written as%
\begin{align}
&  A^{\left\{  p_{1},p_{2},\cdots,p_{n-3}\right\}  ,0,0}\nonumber\\
&  =\prod_{\sigma=1}^{n-3}\left[  \sum_{\sum_{j=1}^{n-2-\sigma}J_{j}^{\sigma
}=\mathcal{P}_{\sigma}}\left(  \mathcal{P}_{\sigma}!\prod_{j=1}^{n-2-\sigma
}\frac{\left(  k_{j+\sigma+1}^{T_{\sigma}}\right)  ^{J_{j}^{\sigma}}}%
{J_{j}^{\sigma}!}\right)  \right] \nonumber\\
&  \times\int_{0}^{1}dz_{n-2}\int_{0}^{1}dz_{n-3}\cdots\int_{0}^{1}dz_{3}%
\int_{0}^{1}dz_{2}\nonumber\\
&  \times z_{2}^{k_{12}}z_{3}^{k_{123}+1-J_{1}^{1}}z_{4}^{k_{1234}+2-J_{1}%
^{1}-\left(  J_{1}^{2}+J_{2}^{1}\right)  }\cdots z_{n-2}^{k_{1,\cdots
,n-2}+\left(  n-4\right)  -J_{1}^{1}-\left(  J_{1}^{2}+J_{2}^{1}\right)
-\cdots-\left(  J_{1}^{n-4}+\cdots+J_{n-4}^{1}\right)  }\nonumber\\
&  \times\left(  1-z_{2}\right)  ^{k_{23}-J_{1}^{1}}\left(  1-z_{2}%
z_{3}\right)  ^{k_{24}-\left(  J_{1}^{2}+J_{2}^{1}\right)  }\cdots\left(
1-z_{2}z_{3}z_{4}\cdots z_{n-2}\right)  ^{k_{2,n-1}-\left(  J_{1}^{n-3}%
+\cdots+J_{n-3}^{1}\right)  }\nonumber\\
&  \times\left(  1-z_{3}\right)  ^{k_{34}}\left(  1-z_{3}z_{4}\right)
^{k_{35}}\cdots\left(  1-z_{3}z_{4}\cdots z_{n-2}\right)  ^{k_{3,n-1}%
}\nonumber\\
&  \vdots\nonumber\\
&  \times\left(  1-z_{n-3}\right)  ^{k_{n-3,n-2}}\left(  1-z_{n-3}%
z_{n-2}\right)  ^{k_{n-3,n-1}}\nonumber\\
&  \times\left(  1-z_{n-2}\right)  ^{k_{n-2,n-1}}.
\end{align}
Let's rearrange the above equation to get a more symmetric form in the following%

\begin{align}
&  A^{\left\{  p_{1},p_{2},\cdots,p_{n-3}\right\}  ,0,0}\nonumber\\
&  =\prod_{\sigma=1}^{n-3}\left[  \sum_{\sum_{j=1}^{n-2-\sigma}J_{j}^{\sigma
}=\mathcal{P}_{\sigma}}\left(  \mathcal{P}_{\sigma}!\prod_{j=1}^{n-2-\sigma
}\frac{\left(  k_{j+\sigma+1}^{T_{\sigma}}\right)  ^{J_{j}^{\sigma}}}%
{J_{j}^{\sigma}!}\right)  \right] \nonumber\\
&  \times\int_{0}^{1}dz_{n-2}\int_{0}^{1}dz_{n-3}\cdots\int_{0}^{1}dz_{3}%
\int_{0}^{1}dz_{2}\cdot\nonumber\\
&  \times z_{2}^{k_{12}}z_{3}^{k_{123}+1-J_{1}^{1}}z_{4}^{k_{1234}+2-J_{1}%
^{1}-\left(  J_{1}^{2}+J_{2}^{1}\right)  }\cdots z_{n-2}^{k_{1,\cdots
,n-2}+\left(  n-4\right)  -J_{1}^{1}-\left(  J_{1}^{2}+J_{2}^{1}\right)
-\cdots-\left(  J_{1}^{n-4}+\cdots+J_{n-4}^{1}\right)  }\nonumber\\
&  \times\left(  1-z_{2}\right)  ^{k_{23}-J_{1}^{1}}\left(  1-z_{3}\right)
^{k_{34}}\left(  1-z_{4}\right)  ^{k_{45}}\cdots\left(  1-z_{n-2}\right)
^{k_{n-2,n-1}}\nonumber\\
&  \times\left(  1-z_{2}z_{3}\right)  ^{k_{24}-\left(  J_{1}^{2}+J_{1}%
^{2}\right)  }\left(  1-z_{2}z_{3}z_{4}\right)  ^{k_{25}-\left(  J_{1}%
^{3}+J_{2}^{2}+J_{3}^{1}\right)  }\cdots\left(  1-z_{2}z_{3}z_{4}\cdots
z_{n-2}\right)  ^{k_{2,n-1}-\left(  J_{1}^{n-3}+\cdots+J_{n-3}^{1}\right)
}\nonumber\\
&  \times\left(  1-z_{3}z_{4}\right)  ^{k_{35}}\cdots\left(  1-z_{3}%
z_{4}\cdots z_{n-2}\right)  ^{k_{3,n-1}}\nonumber\\
&  \vdots\nonumber\\
&  \times\left(  1-z_{n-3}z_{n-2}\right)  ^{k_{n-3,n-1}}.
\end{align}
Then we expand the crossing terms to get%
\begin{align}
&  A^{\left\{  p_{1},p_{2},\cdots,p_{n-3}\right\}  ,0,0}\nonumber\\
&  =\prod_{\sigma=1}^{n-3}\left[  \sum_{\sum_{j=1}^{n-2-\sigma}J_{j}^{\sigma
}=\mathcal{P}_{\sigma}}\left(  \mathcal{P}_{\sigma}!\prod_{j=1}^{n-2-\sigma
}\frac{\left(  k_{j+\sigma+1}^{T_{\sigma}}\right)  ^{J_{j}^{\sigma}}}%
{J_{j}^{\sigma}!}\right)  \right] \nonumber\\
&  \times\int_{0}^{1}dz_{n-2}\int_{0}^{1}dz_{n-3}\cdots\int_{0}^{1}dz_{3}%
\int_{0}^{1}dz_{2}\nonumber\\
&  \times z_{2}^{k_{12}}z_{3}^{k_{123}+1-J_{1}^{1}}z_{4}^{k_{1234}+2-J_{1}%
^{1}-\left(  J_{1}^{2}+J_{2}^{1}\right)  }\cdots z_{n-2}^{k_{1,\cdots
,n-2}+\left(  n-4\right)  -J_{1}^{1}-\left(  J_{1}^{2}+J_{2}^{1}\right)
-\cdots-\left(  J_{1}^{n-4}+\cdots+J_{n-4}^{1}\right)  }\nonumber\\
&  \times\left(  1-z_{2}\right)  ^{k_{23}-J_{1}^{1}}\left(  1-z_{3}\right)
^{k_{34}}\left(  1-z_{4}\right)  ^{k_{45}}\cdots\left(  1-z_{n-2}\right)
^{k_{n-2,n-1}}\nonumber\\
&  \times\sum_{m_{23}}\frac{\left[  -k_{24}+\left(  J_{1}^{2}+J_{2}%
^{1}\right)  \right]  _{m_{23}}}{m_{23}!}\left(  z_{2}z_{3}\right)  ^{m_{23}%
}\cdots\sum_{m_{24}}\frac{\left[  -k_{2,n-1}+\left(  J_{1}^{n-3}%
+\cdots+J_{n-3}^{1}\right)  \right]  _{m_{2,n-2}}}{m_{2,n-2}!}\left(
z_{2}z_{3}z_{4}\cdots z_{n-2}\right)  ^{m_{2,n-2}}\nonumber\\
&  \times\sum_{m_{34}}\frac{\left(  -k_{35}\right)  _{m_{34}}}{m_{34}!}\left(
z_{3}z_{4}\right)  ^{m_{34}}\cdots\sum_{m_{3,n-2}}\frac{\left(  -k_{3,n-1}%
\right)  _{m_{3,n-2}}}{m_{3,n-2}!}\left(  z_{3}z_{4}\cdots z_{n-2}\right)
^{m_{3,n-2}}\nonumber\\
&  \vdots\nonumber\\
&  \times\sum_{m_{n-3,n-2}}\frac{\left(  -k_{n-3,n-1}\right)  _{m_{3,n-2}}%
}{m_{n-3,n-2}!}\left(  z_{n-3}z_{n-2}\right)  ^{m_{n-3,n-2}}%
\end{align}
where the subscripts of $m_{ij}$ keep record of the first and the last
subscripts of $\left(  z_{i}z_{i+1}\cdots z_{j-1}z_{j}\right)  $ etc.. The
amplitude becomes%
\begin{align}
&  A^{\left\{  p_{1},p_{2},\cdots,p_{n-3}\right\}  ,0,0}\nonumber\\
&  =\prod_{\sigma=1}^{n-3}\left[  \sum_{\sum_{j=1}^{n-2-\sigma}J_{j}^{\sigma
}=\mathcal{P}_{\sigma}}\left(  \mathcal{P}_{\sigma}!\prod_{j=1}^{n-2-\sigma
}\frac{\left(  k_{j+\sigma+1}^{T_{\sigma}}\right)  ^{J_{j}^{\sigma}}}%
{J_{j}^{\sigma}!}\right)  \right] \nonumber\\
&  \times\sum_{m_{23}}\frac{\left[  -k_{24}+\left(  J_{1}^{2}+J_{2}%
^{1}\right)  \right]  _{m_{23}}}{m_{23}!}\cdots\sum_{m_{24}}\frac{\left[
-k_{2,n-1}+\left(  J_{1}^{n-3}+\cdots+J_{n-3}^{1}\right)  \right]
_{m_{2,n-2}}}{m_{2,n-2}!}\nonumber\\
&  \times\sum_{m_{34}}\frac{\left(  -k_{35}\right)  _{m_{34}}}{m_{34}!}%
\cdots\sum_{m_{3,n-2}}\frac{\left(  -k_{3,n-1}\right)  _{m_{3,n-2}}}%
{m_{3,n-2}!}\nonumber\\
&  \vdots\nonumber\\
&  \times\sum_{m_{n-3,n-2}}\frac{\left(  -k_{n-3,n-1}\right)  _{m_{3,n-2}}%
}{m_{n-3,n-2}!}\nonumber\\
&  \times\int_{0}^{1}dz_{n-2}\int_{0}^{1}dz_{n-3}\cdots\int_{0}^{1}dz_{3}%
\int_{0}^{1}dz_{2}\cdot\nonumber\\
&  \times z_{2}^{k_{12}+\sum_{i\leq2\leq j}m_{ij}}z_{3}^{k_{123}%
+1-\sum_{i+j\leq2}J_{j}^{i}+\sum_{i\leq3\leq j}m_{ij}}\cdots z_{n-2}%
^{k_{1,\cdots,n-2}+\left(  n-4\right)  -\sum_{i+j\leq n-3}J_{j}^{i}%
+\sum_{i\leq n-2\leq j}m_{ij}}\nonumber\\
&  \times\left(  1-z_{2}\right)  ^{k_{23}-J_{1}^{1}}\left(  1-z_{3}\right)
^{k_{34}}\cdots\left(  1-z_{n-2}\right)  ^{k_{n-2,n-1}}.
\end{align}
After integration, we can write it as%
\begin{align}
&  A^{\left\{  p_{1},p_{2},\cdots,p_{n-3}\right\}  ,0,0}\nonumber\\
&  =\prod_{\sigma=1}^{n-3}\left[  \sum_{\sum_{j=1}^{n-2-\sigma}J_{j}^{\sigma
}=\mathcal{P}_{\sigma}}\left(  \mathcal{P}_{\sigma}!\prod_{j=1}^{n-2-\sigma
}\frac{\left(  k_{j+\sigma+1}^{T_{\sigma}}\right)  ^{J_{j}^{\sigma}}}%
{J_{j}^{\sigma}!}\right)  \right] \nonumber\\
&  \times\sum_{m_{23}}\frac{\left[  -k_{24}+\left(  J_{1}^{2}+J_{2}%
^{1}\right)  \right]  _{m_{23}}}{m_{23}!}\sum_{m_{24}}\frac{\left[
-k_{25}+\left(  J_{1}^{3}+J_{2}^{2}+J_{3}^{1}\right)  \right]  _{m_{24}}%
}{m_{24}!}\cdots\sum_{m_{24}}\frac{\left[  -k_{2,n-1}+\left(  J_{1}%
^{n-3}+\cdots+J_{n-3}^{1}\right)  \right]  _{m_{2,n-2}}}{m_{2,n-2}%
!}\nonumber\\
&  \times\sum_{m_{34}}\frac{\left(  -k_{35}\right)  _{m_{34}}}{m_{34}!}%
\cdots\sum_{m_{3,n-2}}\frac{\left(  -k_{3,n-1}\right)  _{m_{3,n-2}}}%
{m_{3,n-2}!}\nonumber\\
&  \vdots\nonumber\\
&  \times\sum_{m_{n-3,n-2}}\frac{\left(  -k_{n-3,n-1}\right)  _{m_{3,n-2}}%
}{m_{n-3,n-2}!}\nonumber\\
&  \times\frac{\Gamma\left(  k_{12}+1+\sum_{i\leq2\leq j}m_{ij}\right)
\Gamma\left(  k_{23}+1-J_{1}^{1}\right)  }{\Gamma\left(  k_{12}+k_{23}%
+2-J_{1}^{1}+\sum_{i\leq2\leq j}m_{ij}\right)  }\nonumber\\
&  \times\frac{\Gamma\left(  k_{123}+2-\sum_{i+j\leq2}J_{j}^{i}+\sum
_{i\leq3\leq j}m_{ij}\right)  \Gamma\left(  k_{34}+1\right)  }{\Gamma\left(
k_{123}+k_{34}+3-\sum_{i+j\leq2}J_{j}^{i}+\sum_{i\leq3\leq j}m_{ij}\right)
}\nonumber\\
&  \vdots\nonumber\\
&  \times\frac{\Gamma\left(  k_{1,\cdots,n-2}+\left(  n-3\right)
-\sum_{i+j\leq n-3}J_{j}^{i}+\sum_{i\leq n-2\leq j}m_{ij}\right)
\Gamma\left(  k_{n-2,n-1}+1\right)  }{\Gamma\left(  k_{1,\cdots,n-2}%
+k_{n-2,n-1}+\left(  n-2\right)  -\sum_{i+j\leq n-3}J_{j}^{i}+\sum_{i\leq
n-2\leq j}m_{ij}\right)  }.
\end{align}

Now we choose to work on the Regge regime defined by%
\begin{equation}
k_{1,\cdots,n-2}\sim s,k_{1,\cdots,n-2}+k_{n-2,n-1}\sim t.
\end{equation}
In this regime, the RSSA can be approximated as%
\begin{align}
&  A^{\left\{  p_{1},p_{2},\cdots,p_{n-3}\right\}  ,0,0}\nonumber\\
&  \sim\prod_{\sigma=1}^{n-3}\left[  \sum_{\sum_{j=1}^{n-2-\sigma}%
J_{j}^{\sigma}=\mathcal{P}_{\sigma}}\left(  \mathcal{P}_{\sigma}!\prod
_{j=1}^{n-2-\sigma}\frac{\left(  k_{j+\sigma+1}^{T_{\sigma}}\right)
^{J_{j}^{\sigma}}}{J_{j}^{\sigma}!}\right)  \right] \nonumber\\
&  \times\sum_{m_{23}}\frac{\left[  -k_{24}+\left(  J_{1}^{2}+J_{2}%
^{1}\right)  \right]  _{m_{23}}}{m_{23}!}\sum_{m_{24}}\frac{\left[
-k_{25}+\left(  J_{1}^{3}+J_{2}^{2}+J_{3}^{1}\right)  \right]  _{m_{24}}%
}{m_{24}!}\cdots\sum_{m_{2,n-2}}\frac{\left[  -k_{2,n-1}+\left(  J_{1}%
^{n-3}+\cdots+J_{n-3}^{1}\right)  \right]  _{m_{2,n-2}}}{m_{2,n-2}%
!}\nonumber\\
&  \times\sum_{m_{34}}\frac{\left(  -k_{35}\right)  _{m_{34}}}{m_{34}!}%
\cdots\sum_{m_{3,n-2}}\frac{\left(  -k_{3,n-1}\right)  _{m_{3,n-2}}}%
{m_{3,n-2}!}\nonumber\\
&  \vdots\nonumber\\
&  \times\sum_{m_{n-3,n-2}}\frac{\left(  -k_{n-3,n-1}\right)  _{m_{3,n-2}}%
}{m_{n-3,n-2}!}\nonumber\\
&  \times\frac{\Gamma\left(  k_{12}+1+\sum_{i\leq2\leq j}m_{ij}\right)
\Gamma\left(  k_{23}+1-J_{1}^{1}\right)  }{\Gamma\left(  k_{12}+k_{23}%
+2-J_{1}^{1}+\sum_{i\leq2\leq j}m_{ij}\right)  }\nonumber\\
&  \times\frac{\Gamma\left(  k_{123}+2-\sum_{i+j\leq2}J_{j}^{i}+\sum
_{i\leq3\leq j}m_{ij}\right)  \Gamma\left(  k_{34}+1\right)  }{\Gamma\left(
k_{123}+k_{34}+3-\sum_{i+j\leq2}J_{j}^{i}+\sum_{i\leq3\leq j}m_{ij}\right)
}\nonumber\\
&  \vdots\nonumber\\
&  \times\frac{\left(  k_{1,\cdots,n-2}\right)  ^{-\sum_{i+j\leq n-3}J_{j}%
^{i}+\sum_{i\leq n-2\leq j}m_{ij}}\Gamma\left(  k_{1,\cdots,n-2}+\left(
n-3\right)  \right)  \Gamma\left(  k_{n-2,n-1}+1\right)  }{\left(
k_{1,\cdots,n-2}+k_{n-2,n-1}+\left(  n-2\right)  \right)  _{-\sum_{i+j\leq
n-3}J_{j}^{i}+\sum_{i\leq n-2\leq j}m_{ij}}\Gamma\left(  k_{1,\cdots
,n-2}+k_{n-2,n-1}+\left(  n-2\right)  \right)  }.
\end{align}
To get the leading order in $k_{1,\cdots,n-2}\symbol{126}$ $s$, we take%
\begin{equation}
J_{j}^{i}=0,\text{ (for all }i+j\leq n-3\text{)}%
\end{equation}
or%
\begin{align}
J_{1}^{1}  &  =J_{2}^{1}=\cdots=J_{n-4}^{1}=0,\nonumber\\
J_{1}^{2}  &  =\cdots=J_{n-5}^{2}=0,\nonumber\\
J_{1}^{r}  &  =\cdots=J_{n-r-3}^{r}=0
\end{align}
which imply%
\begin{equation}
J_{n-3}^{1}=N+p_{1},J_{n-4}^{2}=p_{2},\cdots,J_{n-r-2}^{r}=p_{r}.
\end{equation}

\bigskip Finally, the leading order term of the amplitude is%
\begin{align}
&  A^{\left\{  p_{1},p_{2},\cdots,p_{r}\right\}  ,0,0}\nonumber\\
&  \sim\prod_{\sigma=1}^{r}\left[  \left(  k_{n-1}^{T_{\sigma}}\right)
^{\mathcal{P}_{\sigma}}\right] \nonumber\\
&  \times\sum_{m_{23}}\frac{\left[  -k_{24}\right]  _{m_{23}}}{m_{23}!}%
\sum_{m_{24}}\frac{\left[  -k_{25}\right]  _{m_{24}}}{m_{24}!}\cdots
\sum_{m_{24}}\frac{\left[  -k_{2,n-1}\right]  _{m_{2,n-2}}}{m_{2,n-2}%
!}\nonumber\\
&  \times\sum_{m_{34}}\frac{\left(  -k_{35}\right)  _{m_{34}}}{m_{34}!}%
\cdots\sum_{m_{3,n-2}}\frac{\left(  -k_{3,n-1}\right)  _{m_{3,n-2}}}%
{m_{3,n-2}!}\nonumber\\
&  \times\sum_{m_{n-3,n-2}}\frac{\left(  -k_{n-3,n-1}\right)  _{m_{3,n-2}}%
}{m_{n-3,n-2}!}\nonumber\\
&  \times\frac{\Gamma\left(  k_{12}+1+\sum_{i\leq2\leq j}m_{ij}\right)
\Gamma\left(  k_{23}+1\right)  }{\Gamma\left(  k_{12}+k_{23}+2+\sum
_{i\leq2\leq j}m_{ij}\right)  }\nonumber\\
&  \times\frac{\Gamma\left(  k_{123}+2+\sum_{i\leq3\leq j}m_{ij}\right)
\Gamma\left(  k_{34}+1\right)  }{\Gamma\left(  k_{123}+k_{34}+3-\sum
_{i+j\leq2}J_{j}^{i}+\sum_{i\leq3\leq j}m_{ij}\right)  }\nonumber\\
&  \vdots\nonumber\\
&  \times\frac{\left(  k_{1,\cdots,n-2}\right)  ^{\sum_{i\leq n-2\leq j}%
m_{ij}}\Gamma\left(  k_{1,\cdots,n-2}+\left(  n-3\right)  \right)
\Gamma\left(  k_{n-2,n-1}+1\right)  }{\left(  k_{1,\cdots,n-2}+k_{n-2,n-1}%
+\left(  n-2\right)  \right)  _{\sum_{i\leq n-2\leq j}m_{ij}}\Gamma\left(
k_{1,\cdots,n-2}+k_{n-2,n-1}+\left(  n-2\right)  \right)  }\nonumber\\
&  =\prod_{\sigma=1}^{r}\left[  \left(  k_{n-1}^{T_{\sigma}}\right)
^{\mathcal{P}_{\sigma}}\right]  \times(\text{factors independent of }J_{q}%
^{r}\text{'}s\text{ }).
\end{align}
The ratios of the amplitudes are
\begin{align}
\frac{A^{\left\{  p_{1},p_{2},\cdots,p_{r}\right\}  ,0,0}}{A^{\left\{
0,0,\cdots,0\right\}  ,0,0}}  &  =\left(  k_{n-1}^{T_{1}}\right)  ^{p_{1}%
}\left(  k_{n-1}^{T_{2}}\right)  ^{p_{2}}\cdots\left(  k_{n-1}^{T_{r}}\right)
^{p_{r}},\nonumber\\
&  =\left(  \Omega_{2}^{n-1}\right)  ^{p_{1}}\left(  \Omega_{3}^{n-1}\right)
^{p_{2}}\cdots\left(  \Omega_{r+1}^{n-1}\right)  ^{p_{r}},\nonumber\\
&  =\left(  \omega_{1}\right)  ^{p_{1}}\left(  \omega_{2}\right)  ^{p_{2}%
}\cdots\left(  \omega_{r}\right)  ^{p_{r}},
\end{align}
which is the same as Eq.(\ref{100}) with $m=q=0$.%

%TCIMACRO{\TeXButton{equation number}{\setcounter{equation}{0}
%\renewcommand{\theequation}{\arabic{section}.\arabic{equation}}}}%
%BeginExpansion
\setcounter{equation}{0}
\renewcommand{\theequation}{\arabic{section}.\arabic{equation}}%
%EndExpansion

\section{Conclusion}

In this paper, we first give a review with detailed calculations of ratios
among HSSA at each fixed mass level to demonstrate the stringy scaling
behavior in the hard scattering limit. We then extend the calculations and
discover a similar stringy scaling behavior for a class of $n$-point RSSA. The
number of independent kinematics variables of these RSSA is found to be
reduced by dim$\mathcal{M}$, similar to those of the HSSA.

These stringy scaling behaviors are reminiscent of deep inelastic scattering
of electron and proton where the two structure functions $W_{1}(Q^{2},\nu)$
and $W_{2}(Q^{2},\nu)$ scale, and become not functions of $2$ kinematics
variables $Q^{2}$ and $\nu$ independently but only of their ratio $Q^{2}/\nu$.
Thus the number of independent kinematics variables reduces from $2$ to $1$.
Indeed, it is now well-known that the structure functions scale as \cite{bs}%
\begin{equation}
MW_{1}(Q^{2},\nu)\rightarrow F_{1}(x),\text{ \ \ }\nu W_{2}(Q^{2}%
,\nu)\rightarrow F_{2}(x) \label{scaling1}%
\end{equation}
where $x$ is the Bjorken variable and $M$ is the proton mass. Moreover, due to
the spin-$\frac{1}{2}$ assumption of quark, Callan and Gross derived the
following relation \cite{cg}%
\begin{equation}
2xF_{1}(x)=F_{2}(x). \label{scaling2}%
\end{equation}

Both of these scaling behaviors, the reduction of the number of kinematics
variables in Eq.(\ref{scaling1}) and the number of structure functions in
Eq.(\ref{scaling2}) in the hard scattering limit of quark-parton model in QCD
seem to persist in some way in the HSSA and some RSSA of string theory. We
believe that, comparing to hard QCD scaling, high energy stringy scaling in
general has not been well studied yet in the literature \cite{oleg}. More new
phenomena of stringy scaling remain to be uncovered.

\begin{acknowledgments}
This work is supported in part by the National Science and Technology Council
(NSTC) and S.T. Yau center of National Yang Ming Chiao Tung University (NYCU),
Taiwan. We thank H. Kawai and Y. Okawa for givng many valuable comments on
stringy scaling before the publication.
\end{acknowledgments}

%\bibliographystyle{unsrt}
%\bibliography{Review}

\end{document}